\documentclass[sigconf, screen, nonacm]{acmart}
\newcommand{\paperTitle}{Optimizing Machine Learning Inference Queries \\
with Correlative Proxy Models \iffull (Technical Report)
\else
\fi}

\usepackage{anyfontsize}
\usepackage[hyphens]{url}

\definecolor{linkcolor}{HTML}{647382}
\definecolor{citecolor}{HTML}{647382} %
\definecolor{urlcolor}{rgb}{0.4,0.2,0.2}
\definecolor{sqlcolor}{HTML}{965d67}
\definecolor{smtcolor}{HTML}{5d968c}
\definecolor{webblue}{rgb}{0,0,.7}
\definecolor{webgreen}{rgb}{0,.5,0}
\definecolor{webbrown}{rgb}{.6,0,0}

\usepackage{amsmath}
\usepackage[cal=boondoxo]{mathalfa}
\usepackage{subcaption}
\usepackage{endnotes,microtype,xspace,graphicx,fancyvrb,multirow}
\usepackage{supertabular,booktabs}
\usepackage{array,underscore, relsize}
\usepackage{fancyhdr}
\usepackage{enumitem}
\usepackage{balance}
\usepackage{booktabs}
\usepackage{pifont}
\usepackage{listings}
\usepackage{multirow}
\usepackage[scaled]{beramono}
\usepackage{tabularx}
\makeatletter
\newcommand\BeraMonottfamily{%
  \def\fvm@Scale{0.85}
  \fontfamily{fvm}\selectfont
}
\makeatother

\definecolor{mymauve}{rgb}{0.58,0,0.82}

\lstdefinestyle{SQLStyle}{
  language=SQL,
  basicstyle={\small\ttfamily},
  breaklines=true,
  frame=none,
  numbers=none,
  keepspaces=true,
  captionpos=b,
  stringstyle=\color{mymauve},
  keywordstyle=\color{blue},
  commentstyle=\color{dkgreen},
}

\lstdefinestyle{ScriStyle}{
language=SQL,
basicstyle=\BeraMonottfamily\footnotesize, 
keywordstyle=\color{smtcolor}\bfseries,
morekeywords={and, or, not},
aboveskip = 0.05in,
belowskip = 0.05in,
literate = {-}{-}1, 
}

\usepackage{aliascnt}

\usepackage[ruled,linesnumbered,noend]{algorithm2e}
\usepackage{setspace}
\usepackage{enumitem}
\usepackage{hyperref}
\usepackage{dsfont}

\SetAlFnt{\small}
\SetAlCapFnt{\small}
\SetAlCapNameFnt{\small}

\usepackage{semantic} 
\usepackage{algorithmicx}
\usepackage[noend]{algpseudocode}
\usepackage{caption}

\usepackage{stmaryrd}
\usepackage{ltablex}
\usepackage{mathtools}
\usepackage{adjustbox}
\usepackage[capitalize,noabbrev,nameinlink]{cleveref}
\newcommand{\mathsymbol}[1]{${#1}$}

\newcommand{\boldstart}[1]{\noindent \textbf{#1}}

\crefname{lstlisting}{listing}{listings}
\Crefname{lstlisting}{Listing}{Listings}

\fvset{fontsize=\scriptsize,xleftmargin=8pt,numbers=left,numbersep=5pt}

\input{code/fmt}

\def\Snospace~{\S{}}

\numberwithin{equation}{section}

\newif\ifdraft\drafttrue
\newif\ifnotes\notestrue
\ifdraft\else\notesfalse\fi

\input{glyphtounicode}
\newcolumntype{R}[1]{>{\raggedleft\let\newline\\\arraybackslash\hspace{0pt}}p{#1}}

\newcommand{\squishitemize}{
 \begin{list}{$\bullet$}
  { \setlength{\itemsep}{0pt}
     \setlength{\parsep}{3pt}
     \setlength{\topsep}{0pt}
     \setlength{\partopsep}{0pt}
     \setlength{\leftmargin}{1.95em}
     \setlength{\labelwidth}{1.5em}
     \setlength{\labelsep}{0.5em} } }

\newcounter{Lcount}
\newcommand{\squishlist}{
    \begin{list}{\arabic{Lcount}. }
   { \usecounter{Lcount}
        \setlength{\itemsep}{0pt}
        \setlength{\parsep}{3pt}
        \setlength{\topsep}{0pt}
        \setlength{\partopsep}{0pt}
        \setlength{\leftmargin}{2em}
        \setlength{\labelwidth}{1.5em}
        \setlength{\labelsep}{0.5em} } }

\newcommand{\squishend}{\end{list}}

\usepackage{xstring}

\newcommand{\rpm}{\sbox0{$1$}\sbox2{$\scriptstyle\pm$}
  \raise\dimexpr(\ht0-\ht2)/2\relax\box2 }

\newtheorem{definition}{Definition}
\newtheorem{theorem}{Theorem}
\newtheorem{lemma}{Lemma}

\newcommand{\cut}[1]{}
\newcommand{\name}{{\sf CORE}}

\newif\iffull
\fulltrue
\fullfalse

\newcommand\vldbdoi{XX.XX/XXX.XX}
\newcommand\vldbpages{XXX-XXX}
\newcommand\vldbvolume{XX}
\newcommand\vldbissue{X}
\newcommand\vldbyear{20XX}
\newcommand\vldbauthors{\authors}
\newcommand\vldbtitle{\shorttitle}
\newcommand\vldbavailabilityurl{}
\newcommand\vldbpagestyle{plain}

\title{\paperTitle} 

\begin{document}

\newcommand{\mail}[1]{\href{mailto:#1}{#1}}

\author{Zhihui Yang$^1$}
\affiliation{
\institution{Fudan University, Shanghai, China}
}
\email{zhyang14@fudan.edu.cn}

\author{Zuozhi Wang}
\affiliation{
\institution{UC Irvine, CA, USA}
}
\email{zuozhiw@ics.uci.edu}

\author{Yicong Huang}
\affiliation{
\institution{UC Irvine, CA, USA}
}
\email{yicongh1@ics.uci.edu}

\author{Yao Lu}
\affiliation{
\institution{Microsoft Research, WA, USA}
}
\email{luyao@microsoft.com}

\author{Chen Li}
\affiliation{
\institution{UC Irvine, CA, USA}
}
\email{chenli@ics.uci.edu}

\author{X.~Sean Wang}
\affiliation{
\institution{Fudan University, Shanghai, China}
}
\email{xywangcs@fudan.edu.cn}

\begin{abstract}
We consider accelerating machine learning (ML) inference queries on unstructured datasets. Expensive operators such as feature extractors and classifiers are deployed as user-defined functions (UDFs), which are not penetrable with classic query optimization techniques such as predicate push-down. Recent optimization schemes (e.g., Probabilistic Predicates or PP) assume independence among the query predicates, build a proxy model for each predicate offline, and rewrite a new query by injecting these cheap proxy models in the front of the expensive ML UDFs. In such a manner, unlikely inputs that do not satisfy query predicates are filtered early to bypass the ML UDFs. We show that enforcing the independence assumption in this context may result in sub-optimal plans. In this paper, we propose CORE, a query optimizer that better exploits the predicate correlations and accelerates ML inference queries. Our solution builds the proxy models online for a new query and leverages a branch-and-bound search process to reduce the building costs. Results on three real-world text, image and video datasets show that CORE improves the query throughput by up to 63\% compared to PP and up to 80\% compared to running the queries as it is.
\end{abstract}
\maketitle

\pagestyle{\vldbpagestyle}
\begingroup\small\noindent\raggedright\textbf{PVLDB Reference Format:}\\
\vldbauthors. \vldbtitle. PVLDB, \vldbvolume(\vldbissue): \vldbpages, \vldbyear.\\
\href{https://doi.org/\vldbdoi}{doi:\vldbdoi}
\endgroup
\begingroup
\renewcommand\thefootnote{}\footnote{\noindent
This work is licensed under the Creative Commons BY-NC-ND 4.0 International License. Visit \url{https://creativecommons.org/licenses/by-nc-nd/4.0/} to view a copy of this license. For any use beyond those covered by this license, obtain permission by emailing \href{mailto:info@vldb.org}{info@vldb.org}. Copyright is held by the owner/author(s). Publication rights licensed to the VLDB Endowment. \\
\raggedright Proceedings of the VLDB Endowment, Vol. \vldbvolume, No. \vldbissue\ %
ISSN 2150-8097. \\
\href{https://doi.org/\vldbdoi}{doi:\vldbdoi} \\
$^1$ Part of the work was done during a visit to UC Irvine.\\
}\addtocounter{footnote}{-1}\endgroup

\ifdefempty{\vldbavailabilityurl}{}{
\vspace{.3cm}
\begingroup\small\noindent\raggedright\textbf{PVLDB Artifact Availability:}\\
The source code, data, and/or other artifacts have been made available at \url{\vldbavailabilityurl}.
\endgroup
}

\section{Introduction}
\label{sec:introduction}
Modern DBMS systems apply machine learning (ML) inference as user-defined functions (UDFs) for complex analytics over unstructured texts, images, and videos~\cite{cai2019model,hilprecht2019deepdb,kunft2019intermediate, lecun2015deep}. Example models include those extracting user sentiments from product reviews for market analysis~\cite{wang2018rafiki} and those estimating vehicle counts from surveillance videos for traffic planning~\cite{ihaddadene2008real}. Consider the following query, where input tweets are processed by two ML UDFs, namely a geographic tagger ($\mathcal{F}_1$) and a sentiment analyzer ($\mathcal{F}_2$), to generate the predicate columns. These queries enable downstream visualization and statistics, such as analytics of election results.
 
\begin{itemize}
\vspace{0.05in} \item[] \texttt{{\bf SELECT}} ${\bf \mathcal{F}_1}$(\texttt{t}) {\bf AS} \texttt{state}, $\mathcal{F}_2$(\texttt{t}) {\bf AS} \texttt{sentiment}
\vspace{0.05in} \item[] \texttt{{\bf FROM} Tweets} {\bf AS} \texttt{t}
\vspace{0.05in} \item[] \texttt{{\bf WHERE} state = `CA' $\wedge$ sentiment = positive;}
\end{itemize} 

\begin{figure}[t]
\centering 
\includegraphics[width=3.4in]{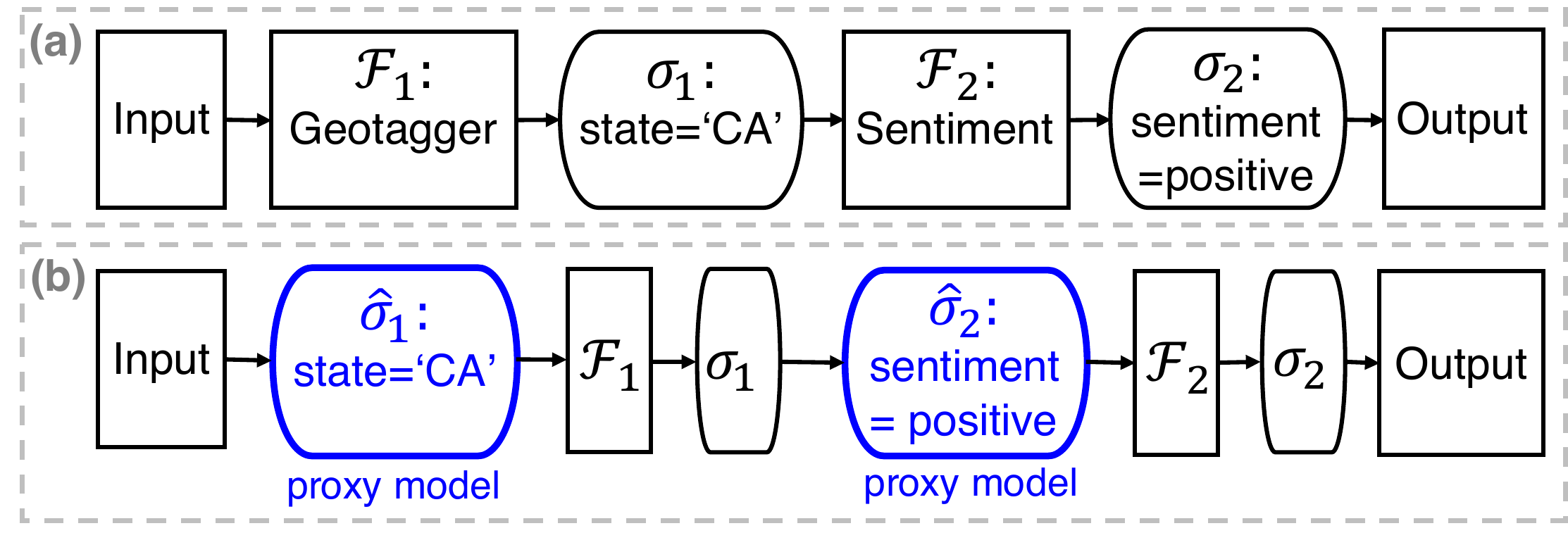}
\caption{(a) An example query plan for tweet analysis. (b) An optimized query plan with proxy models.}
\label{fig:queryplan}
\end{figure}

Figure~\ref{fig:queryplan}(a) demonstrates the plan of the above query, where ${\sigma_1}$ and ${\sigma_2}$ are the predicates \texttt{state = `CA'} and \texttt{sentiment = positive}, respectively. ML queries are costly due to the expensive ML UDFs; improving the efficiency for ML inference has been a recent research focus~\cite{cai2019model,hilprecht2019deepdb,kang2017noscope,krishnan2018deeplens, lu2018accelerating}. In our example, classic query optimization techniques such as predicate push-down cannot help much because $\sigma_1$ and $\sigma_2$ are stuck behind their corresponding ML UDFs regardless of their selectivity.

To optimize such ML inference queries, recent works~\cite{kang2017noscope,lu2018accelerating} propose to rewrite the query and insert a set of light-weight filters in front of the expensive ML UDFs, thus forming a \emph{proxy model}~\cite{viola2001rapid}.  Figure~\ref{fig:queryplan}(b) demonstrates an example plan with two proxy models ${\hat{\sigma}_1}$ and ${\hat{\sigma}_2}$; they quickly discard input records that are unlikely to satisfy the predicates and thus improve the query performance. In~\cite{lu2018accelerating}, a proxy model (i.e., ``Probabilistic Predicate'' or ``PP'') is specific to a predicate \texttt{c}$\phi$\texttt{v}, where \texttt{c} is a predicate column, $\phi$ is a comparison (e.g., $>$ or $=$), and \texttt{v} is a constant value. An independence assumption is made to train filters among different predicates directly using the raw input, regardless of the fact that each may have a different input relation. When ad-hoc queries with multiple predicates arrive, a query optimizer (QO)  rewrites and accelerates the query by assembling individual filters and using them also in an independent manner.
In many applications, query predicates are often correlated. In our example, sentiments may vary in different states -- the sentiment in California can be different from that in Texas.  As Section~\ref{sec:correlation} will show, the QO in~\cite{lu2018accelerating} overestimates the reduction when building the filters on the raw input and thus yields sub-optimal plans for a new query with correlated predicates. 

Inspired by~\cite{lu2018accelerating,kang2017noscope} to optimize ML inference using proxy models, we intend to relax the independence assumption among different predicates. A proxy model hence is specific not only to a predicate but also to its input relation, i.e., prefix $\sigma$'s and $\hat\sigma$'s, as well as parameter choices of prefix $\hat\sigma$'s. In Figure~\ref{fig:queryplan}(b), $\hat\sigma_2$ learns upon filtering the raw input by $\hat\sigma_1\wedge\sigma_1$\footnote{$\mathcal{F}_1$ is a row processor and does not filter as $\sigma_1$ and $\hat\sigma_1$ do.}. Unlike~\cite{lu2018accelerating} that builds a small number of independent filters, it is easy to see that relaxing the independence assumption may result in an untenable number of filters to build by enumerating their order and parameter choices. 

We propose an optimizer called ``CORE'' that better exploits predicate correlations in ML inference. Given an ad-hoc query, CORE builds the proxy models \emph{online} to avoid exhaustive offline filter construction.  We describe a novel technique to accelerate such process at a small overhead (e.g.,  a few percent of the query processing) and a user-specified accuracy target. Extensive experiments for queries over datasets of tweets, images, and videos indicate that CORE improves the ML inference execution costs by up to 63\% compared to~\cite{lu2018accelerating} and up to 80\% compared to running the workload as it is. Various downstream applications, such as interactive data exploration, can benefit from CORE due to a better resource utilization and a faster decision making. 

To summarize, our key contributions are as follows:
\begin{itemize}
\item We show that correlations in predicates may harm the performance of a prior optimization scheme for ML inference~\cite{lu2018accelerating}.
\item We propose CORE to accelerate ML inference and relax the independence assumption enforced by prior work.
Our QO scheme prunes the space of candidate filters to build and incurs only a small computing overhead. 
\item Experiments on real-world ML-inference workloads and datasets show that CORE can achieve significant query-throughput improvements.
\end{itemize}

\subsection{Related Work}
\label{sec:relatedwork}

\vspace{0.05in}
\noindent\textbf{Operator reordering in database optimization.}
\cite{conf/sigmod/HellersteinS93, journals/tods/ChaudhuriS99} studied the problem of reordering select-project-join operators in database systems. \cite{conf/sigmod/BabuMMNW04} studied how to order correlated predicates in streaming systems. It used a greedy algorithm for selection ordering and collected samples at runtime to estimate selectivity. Our query optimization algorithm gives an optimal solution and uses a branch-and-bound search to quickly prune plans in the space of proxy models. \cite{journals/csur/RheinlanderLG17} studied various optimization techniques of complex user-defined functions on map-reduce-style big data systems, such as predicate simplification and UDF semantic inference. These techniques were orthogonal to our solution. Sampling-based approximate query processing techniques~\cite{conf/sigmod/ChaudhuriDK17} provided approximate answers to queries by running queries on a small sampling subset of data. Our approach provides approximate answers by exploiting the accuracy of ML inference predicates.

\vspace{0.05in}
\noindent\textbf{Proxy models (a.k.a.~cascaded filters) in machine learning.} 
One of the first proxy models~\cite{viola2001rapid} cascaded a sequence of light-weight classifiers to discard background regions of an image to accelerate object detection. Later, proxy models were studied to improve the performance of classification~\cite{murthy2016deep}, detection~\cite{cai2015learning,li2015convolutional}, semantic image segmentation~\cite{li2017not}, and pose estimation~\cite{toshev2014deeppose}. Different from~\cite{cai2015learning,li2015convolutional,li2017not,viola2001rapid} that used a cascade of classifiers to quickly reject sub-regions of an image, our {\name} uses proxy models to reduce the size of records to be processed by ML UDFs. Unlike~\cite{murthy2016deep, toshev2014deeppose} that integrated proxy models into DNN models to improve the performance during the training phase, our {\name} uses proxy models as separate operators to accelerate ML inference.

\vspace{0.05in}
\noindent\textbf{Proxy models in databases.} Recently proxy models have been applied in big-data systems to accelerate ML inference-based analysis tasks~\cite{kang2017noscope,kang2018blazeit, kang2020approximate,wang2017idk, lu2018accelerating,hsieh2018focus,kang2020task}. NoScope~\cite{kang2017noscope} firstly cascaded a cheap specialized model before expensive DNNs to accelerate selection video queries. After it, certain classes of video queries including selection without guarantees~\cite{hsieh2018focus}, selection with statistical guarantees~\cite{kang2020approximate}, aggregation~\cite{kang2018blazeit} and limit queries~\cite{kang2018blazeit} was optimized using proxy models. A general index solution in~\cite{kang2020task} was proposed to accelerate these video queries over the schema induced by the target DNN. Probabilistic predicates (PP's)~\cite{lu2018accelerating} optimized various domain queries by inserting multiple offline-built proxy models before expensive ML UDFs with an assumption of independence between predicates. Different from~\cite{wang2017idk,kang2017noscope,hsieh2018focus,kang2020approximate,kang2018blazeit}, {\tt PP} and our proposed {\name} cascade general proxy models, which are applicable to a variety of domains. {\name} follows this line of work and further relaxes the independence assumption of the predicates.

\section{proxy models}
\label{sec:introIdeas}
We briefly review the background of proxy models and then study the impact of correlations to proxy models.

\subsection{Background}
\vspace{0.05in}\boldstart{Proxy models} have been studied for decades to accelerate ML inference. Jones et al.~\cite{viola2001rapid} cascade weak classifiers as proxy models to speed-up face detection in images. Recently, accelerating ML inference with proxy models has attracted attention in relational big-data systems. We briefly review two related solutions~\cite{kang2017noscope,lu2018accelerating} and refer the readers to their papers for  more details.

\vspace{0.05in}\boldstart{NoScope} {\sf (NS)}~\cite{kang2017noscope} aims to process video queries such as ``finding video frames with vehicles'' and ``finding video frames with pedestrians'' using an object-detector UDF. It builds and applies a proxy model, i.e., a cheaper object detector using shallow Neural Networks (NNs), which has the same semantics as the object-detector UDF.  NoScope has to train for each query predicate and thus has large building costs when the query predicates are ad-hoc or complex.
 
\vspace{0.05in}\boldstart{Probabilistic Predicate} {\sf (PP)}~\cite{lu2018accelerating}, as mentioned earlier, {is} another form of proxy models. Each {\sf PP} is a cheap classifier to predict the likelihood of an input record matching a predicate clause. Easy inputs with a small likelihood will be discarded immediately, while hard inputs will be processed further by subsequent ML UDFs. For ad-hoc queries with complex predicates, a query optimizer assembles multiple {\sf PP}s built offline, and a dynamic programming algorithm is leveraged to achieve a maximum reduction, under the independence assumption in queries. However, this assumption made in {\sf PP} limits its use to broader applications. \emph{Dependency between columns is the rule, rather than the exception, in the real world}~\cite{ilyas2004cords}. In the following, we conduct a controlled experiment to study the impact of correlations to proxy models.

\subsection{Impact of Correlations}\label{sec:correlation}
To better understand the impact of  correlations in processing ML inference queries, we leverage the correlation score provided by CORDS~\cite{ilyas2004cords}. Specifically, let  $d_1$ and $d_2$ be the distinct counts in a pair of columns. The correlation score is computed by a chi-squared test upon a sample of $n$-rows: $${\hat{\kappa}^2=\frac{1}{n(\min(d_1,d_2)-1)}\sum_{i=1}^{d_1}\sum_{j=1}^{d_2}\frac{(n_{ij}-n_{i\cdot}n_{\cdot j})^2}{n_{i\cdot}n_{\cdot j}}},$$  where ${n_{ij}}$ is the frequency of distinct tuple ${i,j}$, and ${n_{i\cdot}}$, ${n_{\cdot j}}$ are the marginal frequency. A larger $\hat{\kappa}^2$ value indicates a stronger correlation between the columns. For example, we can follow CORDS to use a sample of 10K rows and normalize the correlations scores by the maximum number in all the predicate pairs. All other algorithmic details follow the CORDS paper~\cite{ilyas2004cords}.

\begin{figure}[hbt!]
\subfloat[Strongly correlated query $q$.]{\includegraphics[width=1.65in]{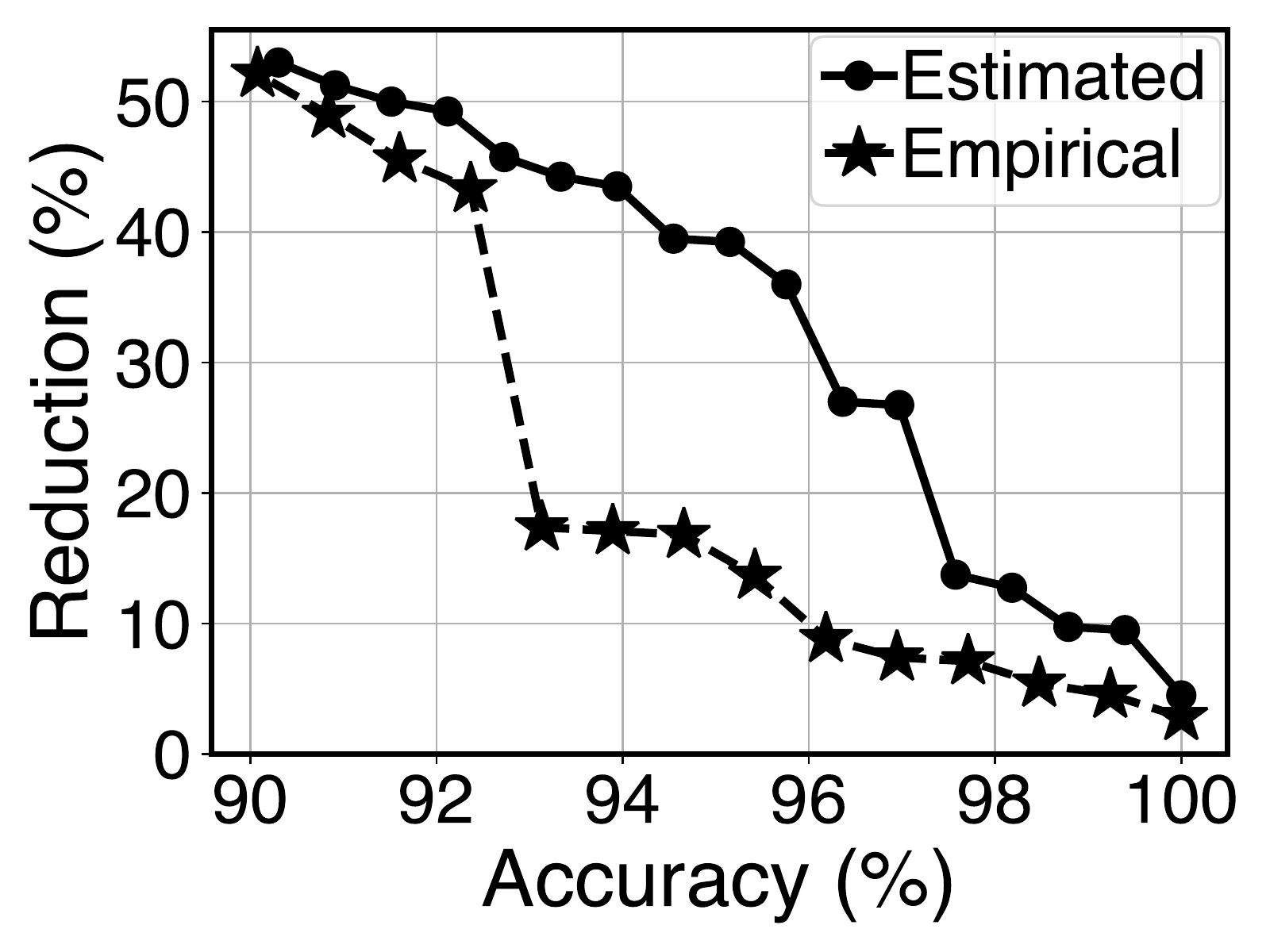}}
\subfloat[Weakly correlated query $q'$.]{\includegraphics[width=1.65in]{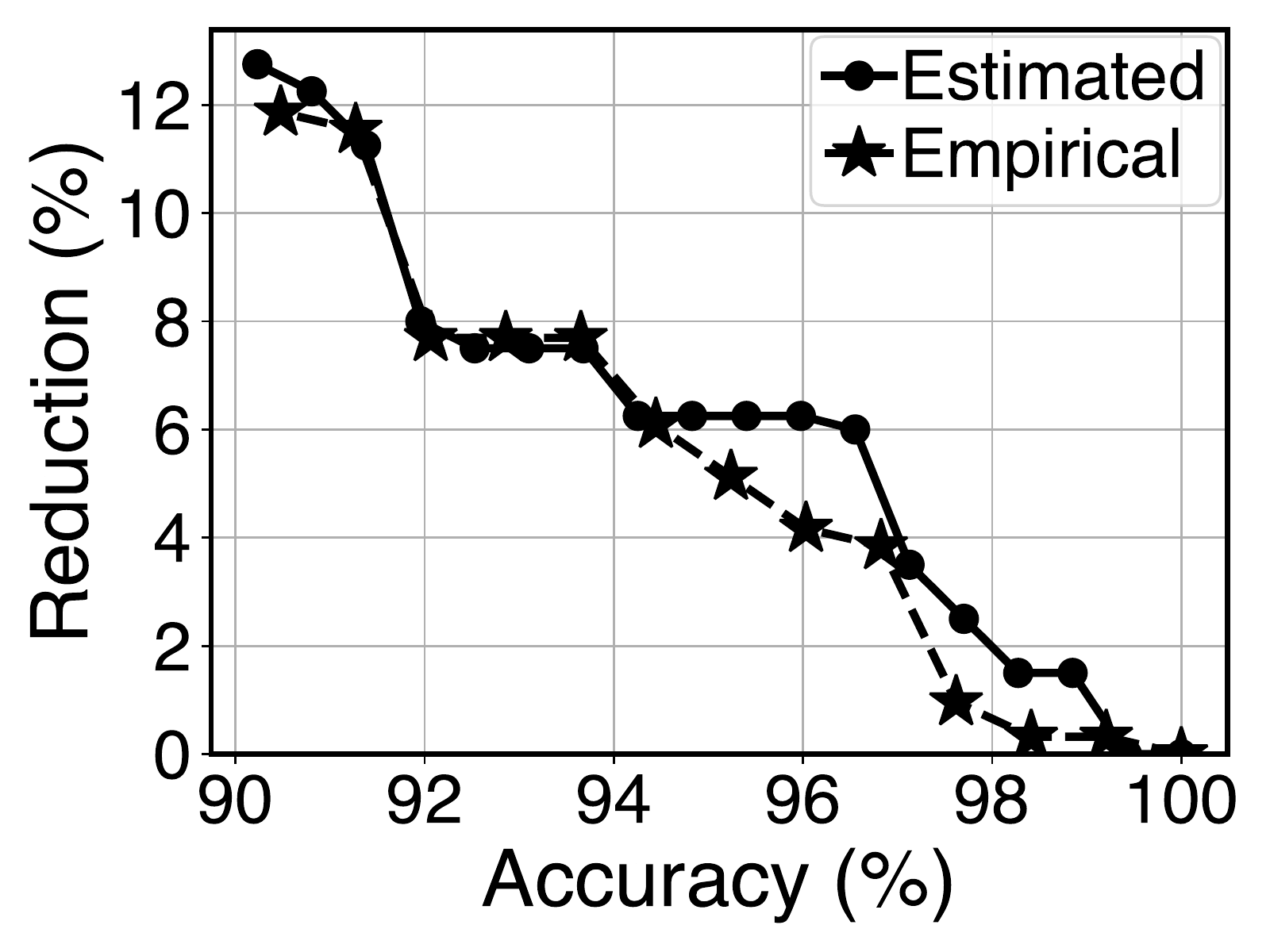}}
\caption{The estimated and empirical accuracy-reduction curves of the \emph{second} {\sf PP} filters in a \emph{strongly} correlated query {\tt ${q}$} and a \emph{weakly} correlated query {\tt ${q'}$}. Correlation results in overestimated reductions offline in {\sf PP}.}
\label{fig:expCorrelation}
\end{figure} 

\vspace{0.05in}
\noindent\textbf{Why correlation matters for {\sf PP}?}
We explain the reason using the Twitter dataset and two queries, $q$ and $q'$, each with two predicates.\footnote{More details of the datasets and queries used in the experiments can be found in Section~\ref{sec:setup}.} The correlation between the $q$ predicates is stronger (2.5 $\times$) than that of the $q'$ predicates. The {\sf PP} filters are trained offline for each predicate without considering the context in which the predicate is applied. The accuracy-reduction curves are estimated during the training as shown in Figure~\ref{fig:expCorrelation}. Two proxy models $\hat{\sigma}_1$ and $\hat{\sigma}_2$ are connected for the predicate $\sigma_1\wedge\sigma_2$. 

When there is a strong correlation between $\sigma_1$ and $\sigma_2$ and $\hat{\sigma}_1$ discards a row that matches $\sigma_1$, it is easy to see that, the discarded row is also likely to match $\sigma_2$ because of the correlation. As shown in Figure~\ref{fig:expCorrelation}, the empirical reduction produced by $\hat{\sigma}_2$ is less than the estimate, because there are fewer input rows for $\sigma_2$ after $\hat{\sigma}_1$. On the contrary, when there is a weak correlation, the reductions are less likely to be overestimated. For example, as shown by ${q}$ with a strong correlation in Figure~\ref{fig:expCorrelation}, when the accuracy is 95\%, the estimated data reduction is 40\%, and the empirical value is 15\%. At the same accuracy, the difference of the reduction ratio for ${q'}$ with a weak correlation is most 2\%. As a result, with strong correlations, {\sf PP} unnecessarily routes more inputs to the expensive ML UDFs and thus yields a lower performance speedup. This example shows that the optimizer in previous work overestimates the reduction of the proxy models built offline, thus yielding suboptimal query plans and less performance improvement for a new query with correlated predicates; this limits the use of {\sf PP}s to broader applications.
\section{CORE Overview}\label{sec:system}
In this section we give an overview of CORE and formally define its optimization problem.

\subsection{System Architecture}
\begin{figure}[hbt!]
\centering
\includegraphics[width=3.3in]{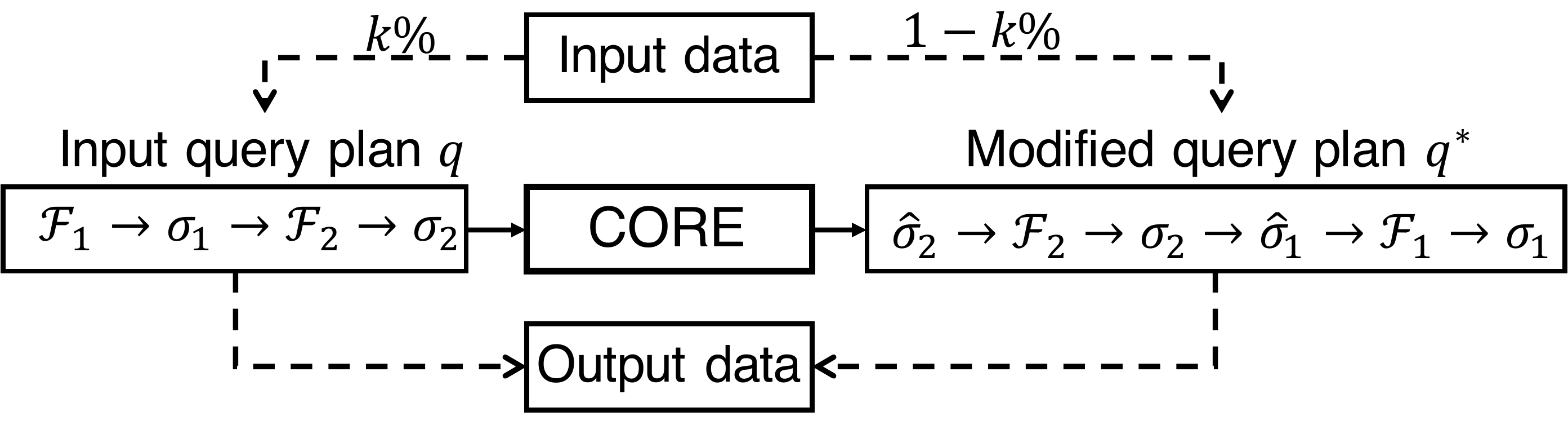}
\caption{Given a query plan $q$, {\name} generates an optimized plan $q^{*}$ by applying proxy models. Part of the input data ($k$\%) is used for  building proxy models, and the remaining data is processed by $q^{*}$.}
\label{fig:sysArch}
\end{figure}

In Figure~\ref{fig:sysArch}, the input of {\name} is a query that includes multiple ML inference UDFs. These UDFs, as seen in the previous section, depict row manipulators; they produce one output row per input row. ML UDFs wrap operations such as feature extraction or classification. {\name} optimizes the input query by building proxy models online and generates a more efficient plan $q^{*}$. We build proxy models for predicates of the form $\texttt{c}\phi\texttt{v}$. Meanwhile, a query can have one or more predicate clauses in conjunction: $\bigwedge{\tt c}\phi{\tt v}$. A small portion of the input data (e.g., $k$\%) is used to build proxy models, and the remaining data is processed by the optimized plan $q^{*}$.

\begin{definition}[]\label{def:proxyModel}
A proxy model ${\hat\sigma}$ is characterized by a tuple $${\{d, \sigma, M, L, R\}},$$
\end{definition}
\noindent where $d$ is an input relation (i.e., applying a sequence of prefix filters on the raw input), and ${\sigma}$ is a target predicate that $\hat\sigma$ aims to improve; ${M}$ is a regression model used by $\hat\sigma$ to produce a scoring function for each input record; ${L}$ is a labeled sample from the input relation $d$ to build $M$; and ${R}$ is a mapping from an accuracy $\alpha$ to a reduction ${r}$. For the example in Figure~\ref{fig:queryplan}(b), $\hat\sigma_1$ is built for the input relation $d_1=\varnothing$ (raw input) and the predicate $\sigma_1:\texttt{state=`CA'}$, while $\hat\sigma_2$ is built for $d_2=(\hat\sigma_1,\sigma_1)$ and $\sigma_2:\texttt{sentiment=positive}$. The mapping $R$ will be explained shortly.

\begin{figure}[hbt!]
\centering 
\includegraphics[width=3.3in]{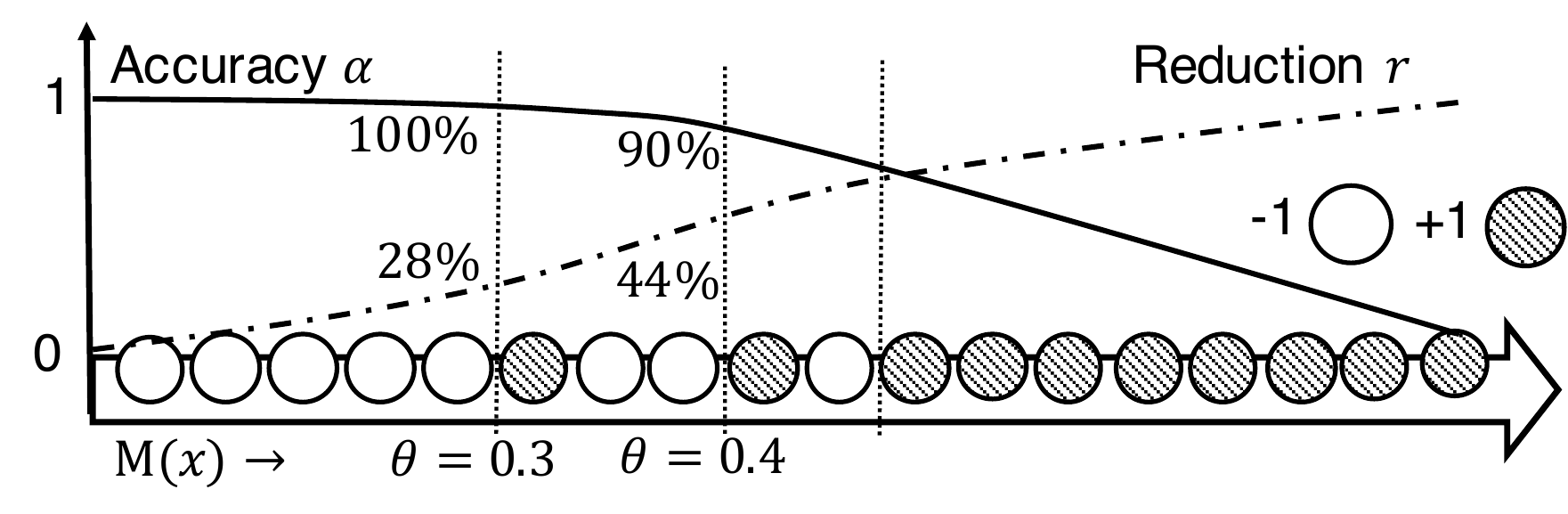} 
\caption{Relationship between an accuracy ${\alpha}$ and a reduction ratio ${r}$ in a proxy model. Records are ranked in the ascending order according to their $M(x)$ score along with $x$-axis. White and dark circles represent records with -1 and +1 labels, respectively.}
\label{fig:buildProxyModel}
\end{figure}

\begin{figure*}[hbt!]
\centering 
\includegraphics[width=7.0in]{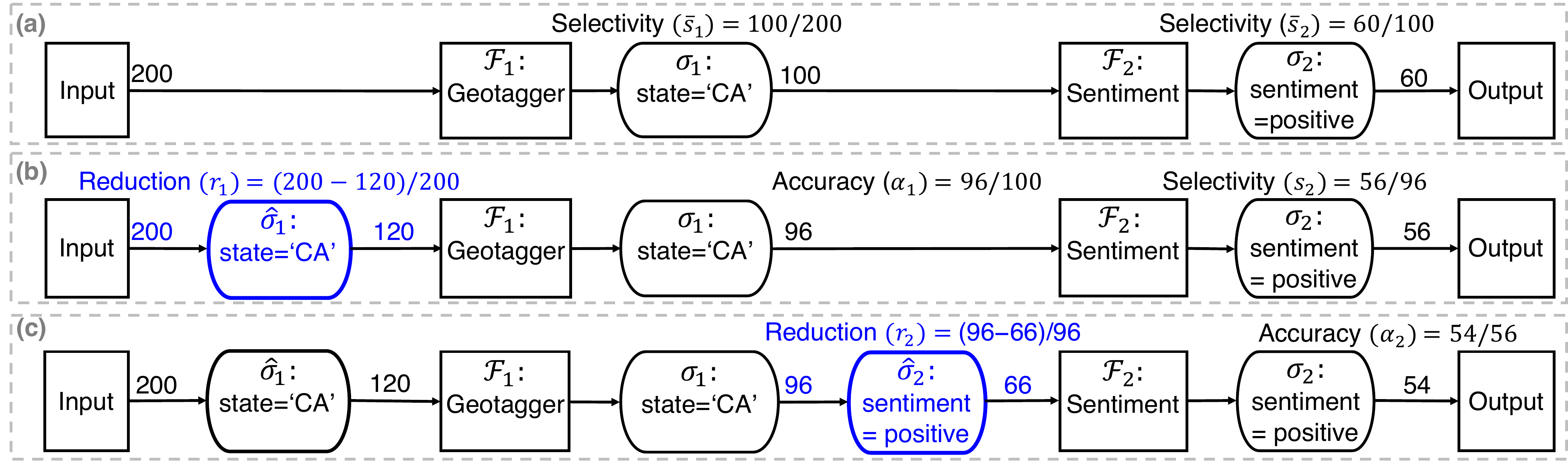}
\vspace{0.001in}
\caption{Step-by-step demonstration of inserting two proxy models to optimize a query. (a) An original query plan; (b) A query plan with $\hat{\sigma}_1$ inserted; (c) A query plan with $\hat{\sigma}_1$ and $\hat{\sigma}_2$ inserted. Each edge depicts the number of passing tweets. Selectivity (i.e., $\bar{s}_i, s_i$), reduction (i.e., $r_i$), and accuracy (i.e., $\alpha_i$) values are illustrated. The overall query accuracy is $\mathcal{A}=54/60$.}
\label{fig:examplePlans}
\end{figure*}

\vspace{0.05in}\boldstart{Building proxy models online} consists of collecting $L$ and then training $M$. We leverage the initial stream of the input data for $L$ (e.g., a few thousand rows). The labeled sample ${L}$ is obtained by applying the filters specified in $d$ upon the raw input and then labeling by predicate $\sigma$. The label is +1 if $\sigma$ is satisfied, and -1 otherwise. Next, we use light-weight regression models such as linear SVMs~\cite{joachims2006training} or shallow NNs~\cite{lecun1990handwritten} to train $M$. Once $M$ is constructed, our developed query optimizer injects $\hat\sigma$ into the query plan right before the corresponding ML UDF that generates the $\sigma$ predicate column (Figure~\ref{fig:queryplan}(b)) for the remaining input records.

Given an input record ${\textbf{x}}$, a proxy model predicts a score ${M(\textbf{x})}$. For example, for linear SVM, ${M(\textbf{x})=\textbf{w}^T\textbf{x}+b}$, where ${\textbf{w}}$ is a weighted matrix and ${b}$ is a bias term. Record ${\textbf{x}}$ will be discarded  if ${M(\textbf{x})<\theta}$ (for a threshold $\theta$), and in this case the record is called a {\em negative example}. It is clear that such early filtering is a trade-off between accuracy and data reduction by setting a proper $\theta$, as shown in Figure~\ref{fig:buildProxyModel}. A higher $\theta$ yields a lower accuracy and a higher data reduction. Note that the mapping between $\alpha$ and $r$ given $\theta$ can be evaluated using a validation set. In the rest of the paper we denote such a relationship as $R$. We can compute it by evaluating $\hat\sigma$ on a validation set from the initial stream of the input records.

\vspace{0.05in}\boldstart{Query optimization by applying proxy models.} We borrow the AQP-style query interface in~\cite{lu2018accelerating}. Specifically, the user issues a query and specifies a global target accuracy ${\mathcal{A}}$ that depicts the level of false negatives of the proxy models in addition to those caused by the UDF. Note that the UDFs themselves produce false positives and negatives and  we do not intend to break the black boxes to improve their accuracy and performance. $\mathcal{A}$ sets the trade-off goals between additional errors and query-processing speedups.  Our QO builds the proxy models, considers their combinations, allocates their accuracy parameters, and injects them into the modified query plan ${q*}$ (Figure~\ref{fig:sysArch}). To reduce the computing overhead and latency of building the proxy models before the input query can be accelerated, the QO reuses intermediate results during the filter construction and prunes candidate plans using a branch-and-bound search. 

\subsection{Formulation of Optimization Problem}
\label{sec:problem}
Given an ML query $q$ with UDFs $\mathcal{F}_1,\ldots,\mathcal{F}_n$, predicate filters $\sigma_1,\ldots,$ $\sigma_n$, and a query-level target accuracy ${\mathcal{A}}$, we aim to build proxy models $\hat\sigma_{1},\ldots,\hat\sigma_{n}$ with their accuracy parameters $\alpha_{1},\ldots,\alpha_{n}$ so that ${\mathcal{A}}$ is met. Let the execution costs of applying $\hat\sigma_i$ and the ML UDF $\mathcal{F}_i$ be ${\hat{c}_i}$ and $c_i$, respectively. The execution cost $C$ of a pair of a proxy model ${\hat\sigma}_i$ and an ML UDF $\mathcal{F}_i$ is
\begin{equation}\label{eq:saveCost}
    C(\hat{\sigma}_i,\alpha_i)=(\prod_{j=1}^{i-1}s_j\cdot\alpha_j)\cdot(\hat{c}_i+(1-r_i)\cdot c_i),
\end{equation}
where $\alpha_i$ is the accuracy of $\hat{\sigma}_i$, $r_i$ is the reduction of $\hat{\sigma}_i$, and $s_i$ is the conditional selectivity of predicate $\sigma_i$ with prior filters $\hat{\sigma}_1,\dots,\hat{\sigma}_{i-1},$ $\sigma_1,\dots,\sigma_{i-1}$.
When building proxy models, their accuracy parameters and $\mathcal{A}$ satisfy
\begin{equation}\label{eq:accuracyRelationship}
    \prod_i\alpha_i\cdot\delta_i=\mathcal{A},
\end{equation}
\noindent where $\delta_i=s_i/\bar{s}_i$, and $\bar{s}_i$ is the conditional selectivity of $\sigma_i$ with prior $\sigma_1,\dots,\sigma_{i-1}$ in the original query. The derivation of Equation~\ref{eq:accuracyRelationship} \iffull\else can be found in our technical report~\cite{ProxymodelTechReport}. 
\fi 

\vspace{0.05in}\boldstart{Example.} We demonstrate the number of passing records by each filter for the example query in Figure~\ref{fig:examplePlans}. According to Equation~\ref{eq:accuracyRelationship},  in Figure~\ref{fig:examplePlans}(a), $\delta_2=s_2/\bar{s}_2$, where $\bar{s}_{2}=60/100$ is the conditional selectivity of predicate ${\tt sentiment = positive}$ with a prior conditional predicate ${\tt state=``CA"}$ (i.e., $\sigma_1$); $s_2=56/96$ is the conditional selectivity of the same predicate with a prior condition $\hat{\sigma}_1\wedge\sigma_1$ in Figure~\ref{fig:examplePlans}(b). Hence, $\delta_{2}=s_{2}/\bar{s}_{2}=(56/96)/(60/100)=0.972$, which measures the changes of the input of ${\sigma}_{2}$ after adding its prefix proxy model $\hat\sigma_{1}$. This proxy model changes the input data size of $\sigma_2$ from 100 to 96 because $\hat{\sigma}_1$ discards 4 tweets satisfying {\tt state=``CA"}. Similarly, $\delta_1=s_1/\bar{s}_1=(100/200)/(100/200)=1$, since $\sigma_1$ is the first filter and there is no prefix proxy model changing the input of $\sigma_1$.

To this end, the right side of Equation~\ref{eq:accuracyRelationship} (i.e., the target accuracy $\mathcal{A}$) is calculated as $\mathcal{A}=54/60=0.9$, which is the percentage of the output of the original query in Figure~\ref{fig:examplePlans}(a) (i.e., 60 tweets) kept by its optimized plan in Figure~\ref{fig:examplePlans}(c) (i.e., 54 tweets). We then consider the left side of Equation~\ref{eq:accuracyRelationship}. For each proxy model $\hat{\sigma}_i$, $\alpha_i$ is the percentage of the output by $\sigma_i$ kept by $\hat{\sigma}_i\wedge\sigma_i$. In Figure~\ref{fig:examplePlans}(b), $\alpha_{1}=96/100=0.96$, as $\hat{\sigma}_1\wedge\sigma_1$ keeps 96 tweets in Figure~\ref{fig:examplePlans}(b) and $\sigma_1$ keeps 100 tweets in Figure~\ref{fig:examplePlans}(a). Similarly, $\alpha_2=54/56=0.964$. As mentioned before, $\delta_1=1$ and $\delta_2=0.972$. Both of them measure the input relation changes for $\sigma_1$ and $\sigma_2$ respectively when applying proxy models. Finally, we have $\alpha_1\cdot\delta_1\cdot\alpha_2\cdot\delta_2=0.9=\mathcal{A}$. In general, relaxing the independence assumption among different predicates results in introducing a input relation change factor $\delta$ caused by its prefix proxy model. For simplicity, we use $\alpha_i$ to refer $\alpha_i\cdot\delta_i$ later on.

\vspace{0.1in}\noindent\boldstart{Problem Statement.}  Let $\pi$ be an order of the ML UDFs and predicate filters. Let $\hat\sigma_{\pi_i}$ denote the $\pi_i$-th proxy model. Our QO  finds the following optimal query plan in the order space $\pi\in\mathbb{H}$ and the accuracy space ${\mathbb{A}}$:
\begin{equation}\label{eq:problem}
    \arg \min_{\pi\in\mathbb{H},\alpha\in\mathbb{A}}\sum_{i}C(\hat\sigma_{\pi_i}, \alpha_{\pi_i}), s.t. \prod_{i} \alpha_{\pi_i}={\mathcal{A}}.
\end{equation}

Finding an optimal order ${\pi}$ of $\hat\sigma$ and allocating their parameter ${\alpha}$\iffull\else simultaneously is NP-hard. Its full proof can be found in the Appendix in our technical report~\cite{ProxymodelTechReport}.
\fi 
Since both $r$ and $s$ depend on $d$ and the input relation of $\hat\sigma$ (i.e., prefix $\sigma$, $\hat\sigma$, and $\alpha$ choices), building $\hat\sigma$ offline by enumerating possible $d$ incurs large computing costs.  We seek a solution such that each $\hat\sigma$ is built on-the-fly on a materialized sample $L$ of its input relation $d$. A main challenge is that, given the accuracy target, how to efficiently build $\hat\sigma$ with a small computing overhead with taking its input relation into account. We describe our solution to find an optimal set of accuracy parameters ${\alpha}\in\mathbb{A}$ given an order $\pi$ in Section~\ref{sec:allocateaccuracy}, and study how to find an optimal order ${\pi}\in\mathbb{H}$ in Section~\ref{sec:reorder}. Both sub-problems exhibit unique structures that can be leveraged for acceleration. Table~\ref{t:param} summarizes the notations used in the paper.

\begin{table}[htbp]
\footnotesize
\centering
\begin{tabular}{c|l}
Notation & Meaning \\\hline
$\sigma$ & A filter predicate after an ML UDF.\\
\hline
$\hat\sigma$ & A cheap proxy model that has the same semantics as $\sigma$. \\
\hline
{${d}$} & The input relation of a proxy model ${\hat{\sigma}}$.\\
\hline
$L$, $M$, ${R}$ & The labeled sample, trained classifier, and accuracy-reduction curve\\
& for a proxy model, respectively.\\
\hline
${\alpha, r}$ & A proxy model's accuracy and the achieved reduction ratio.\\
\hline
${q, \mathcal{A}}$ & A query and a query-level target accuracy specified by a user.\\
\hline
$s_i$ & \multirow{2}{*}{\shortstack[l]{The selectivity of $\sigma_{i}$ on the condition of prefix $\hat{\sigma}_1,\dots,\hat{\sigma}_{i-1}$ and $\sigma_1,$\\ $\dots,\sigma_{i-1}$, i.e., $\sigma_i|(\hat{\sigma}_1,\dots,\hat{\sigma}_{i-1},\sigma_1,\dots,\sigma_{i-1})$.}}\\
&\\
\hline
{${\hat{c_i},c}$}& {The execution cost for ${\hat{\sigma}}$ and an ML UDF ${\mathcal{F}}$.}\\
\hline
$\pi$ & An order of proxy models.\\
\hline
${C_i^l,C_i^u}$ & Lower and upper bounds of execution cost for a pair ($\hat{\sigma}_i$, $\mathcal{F}_i$).\\
\hline
\end{tabular}
\caption{Notations used in this paper.}
\label{t:param}
\end{table}  

\section{CORE: Accuracy Allocation}
\label{sec:allocateaccuracy}
In this section, we present an efficient algorithm in CORE for deriving an optimal accuracy allocation $\alpha_{\pi_1},\ldots,\alpha_{\pi_n}$ among different ${\hat\sigma_{\pi_i}}$ for a given order $\pi$ to achieve a minimum cost $\sum_i C(\hat\sigma_{\pi_i},\alpha_{\pi_i})$. 

\subsection{A Basic Approach and its Challenge}
One approach to allocating the accuracy is as follows. We first discretize ${\mathbb{A}}$ with a fixed step size. For each candidate ${\alpha_{\pi_i}}$ satisfying ${\prod_i\alpha_{\pi_i}\ge\mathcal{A}}$, we build a proxy model in the order of $\pi$. We obtain a labeled sample given its input relation, train a classifier, and derive reduction as mentioned in Section~\ref{sec:system}. After building $\hat{\sigma}_{\pi_i}$, we compute its cost using Equation~\ref{eq:saveCost}, and find an optimal $\alpha$ for a minimal cost. A main challenge is that building  proxy models online is time-consuming for two reasons. (i) There are an exponential number of candidate $\hat{\sigma}_{\pi_i}$s. (ii) For each proxy model, generating an labeled sample and training a classifier can be computationally costly.  

To solve this problem, we present Algorithm~\ref{alg:accuracy-allocation}, which accelerates the construction given input relations specified in $\pi$ by reusing previously materialized samples and trained models. Next we will present the details of the algorithm.

\begin{algorithm}[hbt!]
\caption{Accuracy allocation}\label{alg:accuracy-allocation}
\begin{algorithmic}[1]
\Procedure{Accuracy\_Allocation}{${\pi,\mathcal{A}}$}
\State $L'_{\pi_0}\leftarrow$ raw input;
\State \textbf{for} {$\alpha={\langle\alpha_{\pi_1},\dots,\alpha_{\pi_n}\rangle}$ in discretized ${\mathbb{A}}$}, s.t.~$\prod_i\alpha_{\pi_i}=\mathcal{A}$:\label{alg:aa5}
\State \hspace{0.15in} \textbf{for} $i\in \lbrace1,\dots,n\rbrace$:\label{alg:aa6}
\State \hspace{0.3in} \textbf{if} ${L'_{\pi_i}}$ is not materialized:\label{alg:aa7-if}
\State \hspace{0.45in} ${L'_{\pi_i}}\leftarrow$ Apply $\sigma_{\pi_i}$ on $L'_{\pi_{i-1}}$;\label{alg:aa7-if-body}
\State \hspace{0.3in} $L_{\pi_i}\leftarrow$ Apply $\hat\sigma_{\pi_1},\dots,\hat\sigma_{\pi_{i-1}}$ on $L'_{\pi_i}$ with $\alpha$; \label{alg:aa7} 
\State \hspace{0.3in} Reuse $\hat\sigma^{*}_{\pi_i}$ if ${\epsilon}$\emph{-approx} on $L_{\pi_i}$ else retrain;\label{alg:aa8}
\State \hspace{0.3in} Compute ${C(\hat\sigma_{\pi_i}, \alpha_{\pi_i})}$;\label{alg:aa9}
\State \hspace{0.15in} Compute cost ${\sum_i C(\hat\sigma_{\pi_i},\alpha_{\pi_i})}$;\label{alg:aa10}
\State Pick ${\alpha^{*}}$ in $\mathbb{A}$ with a minimum cost;
\State Retrain $\hat\sigma_{\pi_1},\dots,\hat\sigma_{\pi_n}$ with $\alpha^{*}$;
\State \textbf{return} $\hat\sigma_{\pi_1},\dots,\hat\sigma_{\pi_n}$ and $\alpha^{*}_{\pi_1},\dots,\alpha^{*}_{\pi_n}$.
\EndProcedure
\end{algorithmic}
\end{algorithm}

\subsection{Search Framework}
As shown in Lemma~\ref{theorem:saveCostNonConvex}, the objective function (the cost $\sum_i C(\hat\sigma_{\pi_i},\alpha_{\pi_i})$ subject to $\alpha$) is non-convex, which means there could be multiple locally optimal solutions.  In order to find a globally optimal solution, we use an exhaustive search framework in the algorithm (lines${~\ref{alg:aa5}\sim\ref{alg:aa6}}$). If a locally optimal solution is acceptable by the user, the algorithm can be easily extended to other search frameworks, such as hill climbing, by replacing lines${~\ref{alg:aa5}\sim\ref{alg:aa6}}$.

\begin{lemma}
\label{theorem:saveCostNonConvex}
There exists a non-convex function of the cost $\sum_i C(\hat{\sigma}_i,\alpha_i),$ $s.t. \prod_i\alpha_i=\mathcal{A}$.
\end{lemma}
\begin{proof}
The cost of applying each proxy model before its corresponding ML UDF could be any non-decreasing function over its accuracy, as the reduction decreases with the increase of accuracy.
shown in~\cite{lu2018accelerating}. We prove the lemma by constructing the following example with $n=2$. $$     C(\hat{\sigma}_1,\alpha_1)=1-(\alpha_1-1)^2, \alpha_1\in[0,1].$$ $$C(\hat{\sigma}_2,\alpha_2)=e^{-(2\mathcal{A}/\alpha_2-1)^3}, \alpha_2\in[0,1].$$ Both $C(\hat{\sigma}_1,\alpha_1)$ and $C(\hat{\sigma}_2,\alpha_2)$ increase monotonically when $\alpha_1\in[0,1]$ and $\alpha_2\in[0,1]$. The cost function $f=\sum C$ is $$e^{-(2x-1)^3}+1-(x-1)^2, x\in[0,1].$$

If the function $f$ is convex on an interval $[0,1]$, by definition~\cite{gradshteyn2014table}, for any two points $x_1$ and $x_2$ in $[0,1]$ and any $\lambda$ where $0<\lambda<1$, $$f(\lambda x_1+(1-\lambda)x_2)\le\lambda f(x_1)+(1-\lambda)f(x_2).$$ However, when $x_1=0.1, x_2=0.5$ and $\lambda=1/2$, $f(\frac{x_1+x_2}{2})=1.17; \frac{f(x_1)+f(x_2)}{2}=1.12$. So $f$ does not satisfy $f(\lambda x_1+(1-\lambda)x_2)\le\lambda f(x_1)+(1-\lambda)f(x_2)$. Thus $f$ is not convex.
\end{proof}

\subsection{Reusing Samples to Reduce Labeling Costs}
\label{sec:reuse-sample}
We first give a theorem about the proxy models, then show how the algorithm leverages the theorem to reuse samples.

\subsubsection{Commutative proxy models.} We note that the order of prefix filters is interchangeable as shown in Theorem~\ref{theorem:reorder}. In Figure~\ref{fig:examplePlans}(b), the 96 output tweets after $\hat{\sigma}_1\wedge\sigma_1$ with $\alpha_1=0.96$ are the same as the output tweets of applying $\hat{\sigma}_1$ with $\alpha_1=0.96$ on the  100 output tweets after $\sigma_1$ in Figure~\ref{fig:examplePlans}(a). That is, with $\alpha_1=0.96$, applying $\sigma_1\wedge\hat\sigma_1$ and applying $\hat\sigma_1\wedge\sigma_1$ have the same results. To prove the theorem, we introduce Lemma~\ref{theorem:reorder0} to prove a base case that a pair of $\hat{\sigma}\wedge\sigma$ are commutative, and Lemma~\ref{theorem:reorder1} to prove an inductive case that two pairs of $\hat{\sigma}\wedge\sigma$ are still commutative with the same prefix filter and the same suffix filter, respectively.

\begin{lemma}
\label{theorem:reorder0}
Given a list of records $L$, a filter $\sigma$, and a proxy model $\hat{\sigma}$ with a parameter $\alpha$, $\sigma$ and $\hat{\sigma}$ with $\alpha$ are commutative, i.e., the results after applying $\hat{\sigma}\wedge\sigma$ are the same as that after applying $\sigma\wedge\hat{\sigma}$. That is,  $\hat{\sigma}\wedge\sigma=\sigma\wedge\hat{\sigma}$.
\end{lemma}

\begin{proof}
According to Definition~\ref{def:proxyModel}, a proxy model $\hat{\sigma}$ is built based on its input relation $d$ and a target predicate. After building $\hat{\sigma}$ and allocating an accuracy $\alpha$, $\hat{\sigma}$ is a selection predicate with fixed values of $\alpha$, $r$, and $M$. When applying $\hat{\sigma}$, any input record cannot change $\hat{\sigma}$. $\hat{\sigma}$ predicts the same output for a record in different orders (e.g., $\sigma\wedge\hat{\sigma}$ and $\hat{\sigma}\wedge\sigma$). For $\sigma\wedge\hat{\sigma}$, an unseen record $x$ for $\hat{\sigma}$ is the one passed by $\sigma$. If $\hat{\sigma}$ passes $x$, then $x$ is in the output of $\sigma\wedge\hat{\sigma}$ and also in the output of $\hat{\sigma}\wedge\sigma$. Otherwise, $x$ is not in their outputs. For $\hat{\sigma}\wedge\sigma$, $\hat{\sigma}$ takes more input records, compared to $\sigma\wedge\hat{\sigma}$. There is no unseen record for $\hat{\sigma}$. As selection predicates are commutative in general, $\sigma$ and $\hat{\sigma}$ with $\alpha$ are commutative.
\end{proof}

\begin{lemma}
\label{theorem:reorder1}
Given a list of records $L$, two filters $\sigma_i$ and $\sigma_j$, and a proxy model $\hat{\sigma}_k$ with a specific parameter $\alpha_k$, we have
\begin{equation}\label{eq:reorder11}
    \hat{\sigma}_k\wedge\sigma_i\wedge\sigma_j=\sigma_i\wedge\hat{\sigma}_k\wedge\sigma_j,
\end{equation}
and
\begin{equation}\label{eq:reorder12}
    \sigma_i\wedge\hat{\sigma}_k\wedge\sigma_j=\sigma_i\wedge\sigma_j\wedge\hat{\sigma}_k.
\end{equation}
\end{lemma}

\begin{proof}
We first prove $\hat{\sigma}_k$ and $\sigma_i$ are commutative with the same suffix $\sigma_j$ (i.e., Equation~\ref{eq:reorder11}). According to Lemma~\ref{theorem:reorder0}, $\hat{\sigma}_k$ and $\sigma_i$ are commutative (i.e., $\hat{\sigma}_k\wedge\sigma_i=\sigma_i\wedge\hat{\sigma}_k$). $\hat{\sigma}_k\wedge\sigma_i$ and $\sigma_i\wedge\hat{\sigma}_k$ followed by the same suffix $\sigma_j$ produce the same outputs, i.e., $\hat{\sigma}_k\wedge\sigma_i\wedge\sigma_j=\sigma_i\wedge\hat{\sigma}_k\wedge\sigma_j$.

Next, we prove $\hat{\sigma}_k$ and $\sigma_j$ are commutative with the same prefix $\sigma_i$ (i.e., Equation~\ref{eq:reorder12}). For $\sigma_i\wedge\hat{\sigma}_k\wedge\sigma_j$ and $\sigma_i\wedge\sigma_j\wedge\hat{\sigma}_k$, the input of $\hat{\sigma}_k\wedge\sigma_j$ and that of $\sigma_j\wedge\hat{\sigma}_k$ are the same because of the same input list of records $L$ and the same prefix $\sigma_i$. Based on Lemma~\ref{theorem:reorder0}, $\sigma_i\wedge\hat{\sigma}_k\wedge\sigma_j=\sigma_i\wedge\sigma_j\wedge\hat{\sigma}_k$.
\end{proof}

\begin{theorem}
\label{theorem:reorder}
Given a sample of records $L$, filters $\sigma_1,\dots,\sigma_n$, and proxy models $\hat{\sigma}_1,\dots,\hat{\sigma}_n$ with specific parameters $\alpha_1,\dots,\alpha_n$, we have:
$$\bigwedge_{i=1}^n(\hat{\sigma}_i\wedge\sigma_i)=(\bigwedge_{i=1}^n\sigma_i)\wedge(\bigwedge_{i=1}^n\hat{\sigma}_i).$$
\end{theorem}
\begin{proof}
We prove the claim by induction.

\vspace{0.05in}\noindent{Base case: $n=1$.} According to Lemma~\ref{theorem:reorder0}, $\hat{\sigma}_1\wedge\sigma_1=\sigma_1\wedge\hat{\sigma}_1$.

\vspace{0.05in}\noindent{Inductive case: $n=m+1$.} Assume that $\bigwedge_{i=1}^m(\hat{\sigma}_i\wedge\sigma_i)=(\bigwedge_{i=1}^m\sigma_i)\wedge(\bigwedge_{i=1}^m\hat{\sigma}_i)$ with $n=m$. Next, we prove that the claim is also true for $n=m+1$.

First, according to the assumption for $n=m$, Expression~\ref{eq:reorder1} equals to Expression~\ref{eq:reorder2}. Next, based on Lemma~\ref{theorem:reorder1}, a proxy model and a filter are commutative when they have the same prefix filters. Expression~\ref{eq:reorder2} equals to Expression~\ref{eq:reorder3}. Additionally, a proxy model and a filter are commutative when they have same suffix filters according to Lemma~\ref{theorem:reorder1}. We exchange $\sigma_{m+1}$ in Expression~\ref{eq:reorder3} with prefix $m$ proxy models in turn using Lemma~\ref{theorem:reorder1}. Therefore, Expression~\ref{eq:reorder3} equals to Expression~\ref{eq:reorder4}.
\begin{align}
    & (\hat{\sigma}_1\wedge\sigma_1\wedge\dots\wedge\hat{\sigma}_m\wedge\sigma_m)\wedge\hat{\sigma}_{m+1}\wedge\sigma_{m+1}\label{eq:reorder1}\\
    = &(\sigma_1\wedge\dots\wedge\sigma_m\wedge\hat{\sigma}_1\wedge\dots\wedge\hat{\sigma}_m)\wedge\hat{\sigma}_{m+1}\wedge\sigma_{m+1}\label{eq:reorder2}\\
    = &(\sigma_1\wedge\dots\wedge\sigma_m\wedge\hat{\sigma}_1\wedge\dots\wedge\hat{\sigma}_m)\wedge\sigma_{m+1}\wedge\hat{\sigma}_{m+1}\label{eq:reorder3}\\
    = &(\sigma_1\wedge\dots\wedge\sigma_m\wedge\sigma_{m+1})\wedge(\hat{\sigma}_1\wedge\dots\wedge\hat{\sigma}_m\wedge\hat{\sigma}_{m+1})\label{eq:reorder4}.
\end{align}
Therefore, $\bigwedge_{i=1}^n(\hat{\sigma}_i\wedge\sigma_i)=(\bigwedge_{i=1}^n\sigma_i)\wedge(\bigwedge_{i=1}^n\hat{\sigma}_i)$.
\end{proof}

\subsubsection{Reusing samples.}
The algorithm improves the performance by reusing early samples (lines~\ref{alg:aa7-if}~to~\ref{alg:aa7}).  $L_{\pi_i}$ is the sampled input to build ${\hat\sigma_{\pi_i}}$ by applying predicate $\sigma_{\pi_i}$ on the input relation $d_{\pi_i}$. In Figure~\ref{fig:examplePlans}(b), the labeled sample $L_2$ for $\hat{\sigma}_2$ has 96 tweets, which are filtered by $\hat{\sigma}_1\wedge\sigma_1$ on the raw input and then labeled using the predicate \texttt{sentiment=positive}. It is easy to see that $L_{\pi_i}$ changes when accuracies assigned to its prefix proxy models (i.e., ${\alpha_{\pi_1},\ldots,\alpha_{\pi_{i-1}}}$) change. For example, in Figure~\ref{fig:examplePlans}(b), ${L_2}$ changes from 97 tweets to 96 tweets when the accuracy parameter of its prefix $\hat{\sigma}_1$ changes from $\alpha_1=0.97$ to $\alpha_1=0.96$.

By leveraging Theorem~\ref{theorem:reorder}, we can improve the performance by materializing samples $L'$ after $\sigma$, and applying $\hat\sigma$ on $L'$ during the search, since common $L'$ can be shared for different $\alpha$ choices. $L_{\pi_i}$ can be obtained by applying $\hat\sigma_{\pi_1},\ldots,\hat\sigma_{\pi_{i-1}}$ on a pre-computed sample $L'_{\pi_i}$ that is computed by applying $\sigma_{\pi_1},\ldots,\sigma_{\pi_{i-1}}$ on the raw input. Lines~\ref{alg:aa7-if}~to~\ref{alg:aa7} illustrate this process of quickly deriving $L$ for each $\alpha$ search. For the proxy model $\hat{\sigma}_2$, we materialize its corresponding sample $L'_2$ containing 100 tweets filtered by $\sigma_1$ in Figure~\ref{fig:examplePlans}(a) to be reused. When $\alpha_1=0.97$, the labeled sample $L_2$ can be obtained by applying prefix $\hat{\sigma}_1$ with $\alpha_1=0.97$ on the 100 materialized  tweets and producing 97 tweets. Similarly, when $\alpha_1$ changes to 0.96 in Figure~\ref{fig:examplePlans}(b), the labeled sample $L_2$ can be obtained by applying $\hat{\sigma}_1$ with $\alpha_1=0.96$ on the already materialized sample $L'_2$ of 100 tweets and producing 96 tweets. This solution is simple but effective, since applying $\hat\sigma$ is cheap and doing so allows us to evaluate each expensive $\mathcal{F}$ and $\sigma$ only once. 

\subsection{Reusing Classifiers to Reduce Training Costs}\label{sec:reuse-PP}
The algorithm adopts a classifier-reusing scheme (line~\ref{alg:aa8}) to avoid repeated training when the prefix proxy models change their accuracy assignments.  Specifically, let $\hat\sigma^{*}$ trained on ${L^{*}}$ with $\alpha$ from a previous iteration (line~\ref{alg:aa5}) be ${\epsilon}$-approximate \cite{agarwal2005geometric} to $\hat\sigma$ trained on ${L}$.  That is:
\begin{equation}\label{eq:approximate}
    (1-\epsilon)\phi^{*}(L^{*})\le\phi^{*}L\le(1+\epsilon)\phi^{*}L^{*},
\end{equation}
where $\phi$ is the objective function of the regressor model used by the proxy model. $\phi$ can be computed using a scoring function, such as F1 score or coreset~\cite{agarwal2005geometric}. Take the F1 scoring function as an example. We efficiently compute $\phi$ by evaluating $\hat\sigma^{*}$ from a previous iteration and measuring its F1 score on its labeled sample ${L^{*}}$ and current ${L}$~\cite{agarwal2005geometric}. $\hat\sigma^{*}$ can be reused if it is ${\epsilon}$-approximate under the current accuracy setting. In Figure~\ref{fig:examplePlans}(b), suppose we want to build the proxy model $\hat{\sigma}_2$ for the predicate {\tt sentiment=positive} on its 96 labeled tweets with prefix $\alpha_1=0.96$. If there is a proxy model $\hat{\sigma}_2^{*}$ trained on 97 tweets with prefix $\alpha_1=0.97$ satisfying Equation~\ref{eq:approximate}, we reuse the classifier in $\hat{\sigma}_2^{*}$ (i.e., $M_2^{*}$) without training a new classifier on the 96 tweets. In Equation~\ref{eq:approximate}, we compute $\phi^{*}(L^{*})$ by evaluating the F1 score of $M_2^{*}$ on the 97 tweets, while $\phi^{*}(L)$ is on the 96 tweets. 

We next discuss how to compute $C(\hat{\sigma}_i,\alpha_i)$ (line~\ref{alg:aa9}). The per-row cost ${\hat{c}}$ for $\hat\sigma$ and ${c}$ for $\mathcal{F}$ can be profiled during training or by counting the FLOPS of the ML model, while $r$ can be obtained from $R$, and $s$ can be measured by applying the prefix filters on a sample of the raw input. Since applying the proxy models is computationally cheap, $C$ can be computed efficiently. In Figure~\ref{fig:examplePlans}, the cost of the ML UDF {\tt Geotagger} is 20ms per tweet in our experiments, while that of the proxy model $\hat{\sigma}_1$ is 0.01ms per tweet. The proxy model $\hat{\sigma}_1$ with $\alpha_1=0.96$ pays the cost of processing 200 tweets and saves the cost of the 80 discarded tweets, which no longer need to be processed by the ML UDF {\tt Geotagger}. Therefore, using Equation~\ref{eq:saveCost}, we have $C(\hat{\sigma}_1,\alpha_1)=\hat{c}_1+(1-r_1)\cdot c_1=0.01+(1-80/200)\cdot20=12.01$.
\section{CORE: Reordering Proxy Models}
\label{sec:reorder}

In this section we study how to reorder proxy models to find an optimal order $\pi\in\mathbb{H}$ to minimize the cost $\sum C$. For different orders, proxy models built on input relations and predicates are different and they have different costs. For instance, in Figure~\ref{fig:examplePlans}(c), for the order {\tt state = ``CA''$\wedge$sentiment = positive}, the proxy model for predicate {\tt state = ``CA''} is built on the original input data. For the order {\tt sentiment = positive$\wedge$state = ``CA''}, the proxy model for the same predicate is built on records satisfying the predicate {\tt sentiment = positive}. Because different orderings affect the input data to the proxy model, these two proxy models have different execution costs for the same ML UDF {\tt Geotagger}.

The number of query plans in ${\mathbb{H}}$ is exponential in terms of the number of UDFs and filters. We construct a search tree to represent them by merging common prefixes of query plans. For example, let \mathsymbol{X}, \mathsymbol{Y}, and \mathsymbol{Z} be three ML UDFs. There are six potential plans in $\mathbb{H}$ (e.g., \mathsymbol{XYZ} and \mathsymbol{XZY}). Figure~\ref{fig:4bbSearchAlg} shows a snippet of the search tree starting from node \mathsymbol{X}, where each tree node represents an ML UDF $\mathcal{F}$ and its corresponding $\hat\sigma$ and $\sigma$. In general, building all proxy models for the plans can be computationally prohibitive. To find an optimal order ${\pi}$ efficiently, we propose a search algorithm based on branch-and-bound~\cite{kohler1974characterization, little1963algorithm} to prune candidate plans.

\subsection{Bounded Cost}
\label{sec:bbInsight}
\begin{figure}[hbt!]
\centering
\includegraphics[width=3.4in]{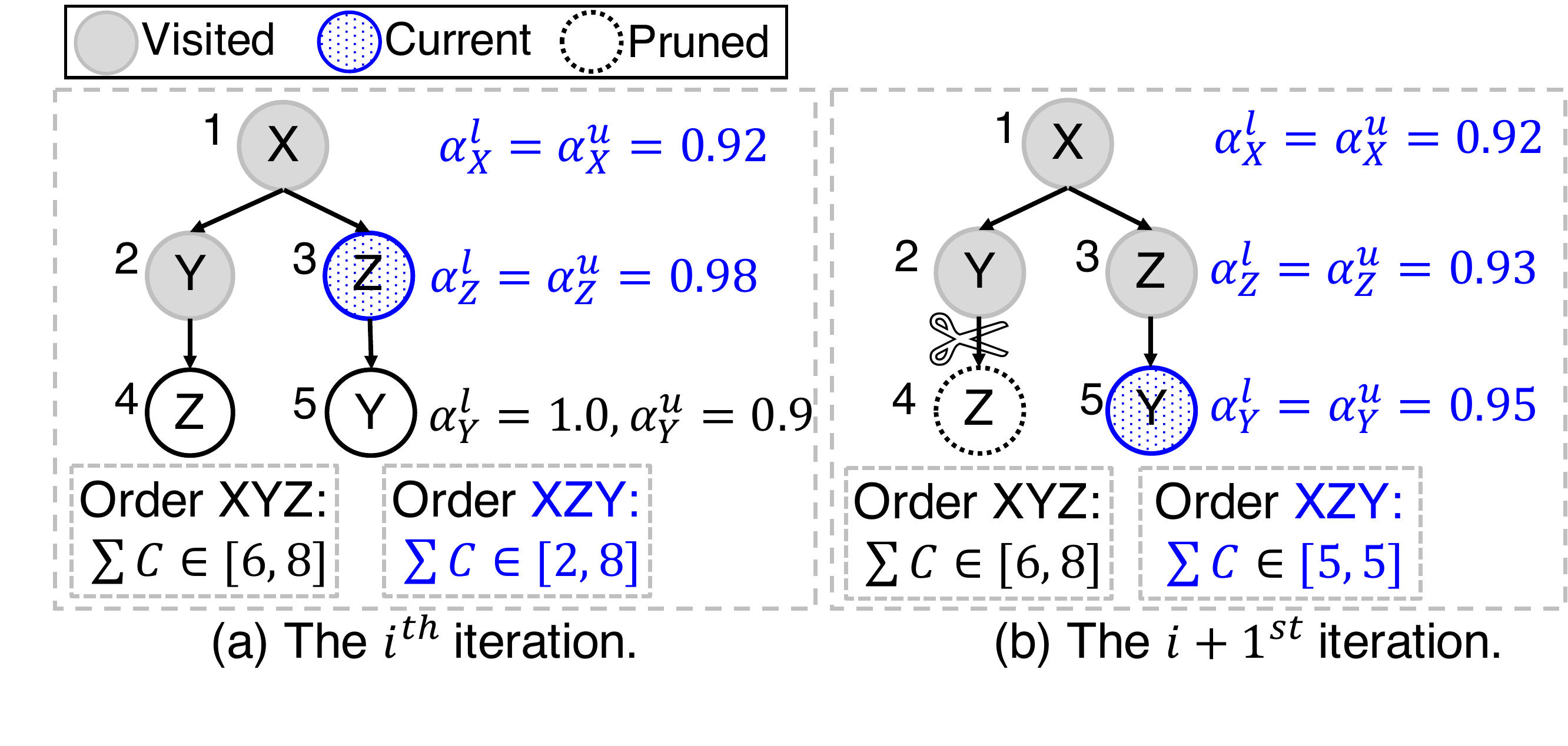}
\caption{Two iterations in branch-and-bound search on a tree starting from node 1 with $\mathcal{A}=0.9$. The blue text is updated information such as accuracies, lower bounds, and upper bounds after calling the function \texttt{update\_node()}.}
\label{fig:4bbSearchAlg}
\end{figure}

For a specific order of proxy models, we can compute a lower bound and an upper bound of the cost $\sum C$. Intuitively, an initial lower bound corresponds to the case when all proxy models discard everything. An initial upper bound corresponds to the case when all proxy models discard nothing. For example, for the order \mathsymbol{XYZ} in Figure~\ref{fig:4bbSearchAlg}, the cost function reaches a lower bound when the first proxy model $\hat{\sigma}_X$ discards all its input records. It reaches an upper bound when all proxy models $\hat{\sigma}_{X}$, $\hat{\sigma}_{Y}$, and $\hat{\sigma}_{Z}$ discard nothing. 

Let $C^l$ and $C^u$ be the lower and upper bounds of the cost for a node, respectively. As shown in Equation~\ref{eq:saveCost}, the cost $C$ of a proxy model $\hat{\sigma}$ is bounded by accuracy $\alpha$, reduction $r$, and selectivity $s$, where (i) $\alpha\in[\mathcal{A},1]$, (ii) $s\in[0,1]$ and (iii) $r\in[0,1]$. $C$ increases when $s$ and $\alpha$ increase and $r$ decreases. To calculate a lower bound of node $t$ at depth $i$ assuming the depth of the root is $0$, we use the minimal value of the accuracy $\alpha^l_i=\mathcal{A}$, the minimal value of the selectivity $s^l_i=0$, and the maximum value of the reduction $r^u_i=1$.  Similarly, to compute an upper bound of $t$, we use the maximum value of the accuracy $\alpha^u_i=1$, the maximum value of the selectivity $s^u_i=1$, and the minimal value of the reduction $r^l_i=0$. Based on the analysis, we present a lower bound and an upper bound of the cost $C$ of a node $t$ in Lemma~\ref{lemma:bound}. Additionally, a lower bound of the cost for a plan is the sum of the lower bound of the cost for each node in the plan, and an upper bound for a plan is the sum of the upper bound for each node in the plan. That is, the bounds of $\sum C$ for a plan are $\sum C^l$ and $\sum C^u$, respectively.

\begin{lemma}[]
\label{lemma:bound}
For a tree node $t$ of depth $i$, a lower bound of its cost $C_t$ is
\begin{equation}\label{eq:lowerBound}
    (\prod_{j=1}^{i-1}s_j^l\cdot\alpha_j^l)\cdot\bigl(\hat{c}_i+(1-r_i^u)\cdot c_i\bigr).
\end{equation}
An upper bound is
\begin{equation}\label{eq:upperBound}
    (\prod_{j=1}^{i-1}s_j^u\cdot\alpha_j^u)\cdot\bigl(\hat{c}_i+(1-r_i^l)\cdot c_i\bigr).
\end{equation}
\end{lemma}

\boldstart{Example.} In Figure~\ref{fig:4bbSearchAlg}, the lower bound of node 1 is the cost of applying a proxy model. $C_X^l=\hat{c}_X$ using Expression~\ref{eq:lowerBound} with $\alpha_X^l=\mathcal{A}$, $s_X^l=0$, and $r_X^u=1$. The upper bound $C_X^u$ is the cost of a proxy model $\hat{c}_X$ plus that of the ML UDF $c_X$ with $\alpha_X^u=1$, $s_X^u=1$, and $r_X^l=0$. For the plan $XYZ$ in Figure~\ref{fig:4bbSearchAlg}, the lower bound of the plan is $C_X^l+C_Y^l+C_Z^l$, and the upper bound is $C_X^u+C_Y^u+C_Z^u$.

\subsection{Branch-and-bound Search}
\label{sec:branchandbound}

We present a general pruning framework in Algorithm~\ref{alg:B-B-pruning}. Its main idea is that the upper and lower bounds can be improved as we collect information during the search process, such as selectivity and reduction. The search builds necessary proxy models and prunes the search tree to reduce the optimization overhead. For each node $t$,  according to Lemma~\ref{lemma:bound}, we initialize the lower and upper bounds of $\hat{\sigma}$ using $C^l$ and $C^u$, respectively   (lines~\ref{alg:bb4}${\sim}$~\ref{alg:bb5}). We then progressively build proxy models (lines~\ref{alg:bb6}${\sim}$\ref{alg:bb11}). For each search step, we find optimal $\alpha$ parameters for $t$ and prefix nodes using Algorithm~\ref{alg:accuracy-allocation}. We compute the cost $\sum C$ of these nodes after using Algorithm~\ref{alg:accuracy-allocation}, and tighten the bounds of costs for $t$'s leaf nodes. The search yields an order $\pi$ that minimizes the overall cost ${\sum_i C(\hat\sigma_{\pi_i},\alpha_{\pi_i})}$. We next explain several specific functions used in the algorithm.

\begin{algorithm}[t]
\caption{QO by branch-and-bound pruning}\label{alg:B-B-pruning}
\begin{algorithmic}[1]
\Procedure{bb\_pruning}{$q$, $\mathcal{A}$}
\State Construct a search tree based on $\mathbb{H}$ from $q$;\label{alg:bb2}
\State $Q$ = \{$q_\pi|\forall\pi\in\mathbb{H}$\}; {\tt visited}=$\varnothing$;\label{alg:bb3}
\State \textbf{for} each node $t$ in the search tree:\label{alg:bb4}
\State \hspace{0.1in} $C^l,C^u\leftarrow$\texttt{initialize}($t$);\label{alg:bb5}
\State \textbf{while} $|Q|>1$:\label{alg:bb6}
\State \hspace{0.1in} $t\leftarrow$ \texttt{pop\_unvisited}($Q$, {\tt visited});\label{alg:bb7}
\State \hspace{0.1in} $\hat{\sigma}^{*},\alpha^{*}\leftarrow$ {\tt accuracy\_allocation}($t, \mathcal{A}$);\label{alg:bb8}
\State \hspace{0.1in} \texttt{update\_{}node($t,\hat{\sigma}^{*},\alpha^{*}$);}\label{alg:bb9}
\State \hspace{0.1in} {\tt visited} = {\tt visited} $\cup \{t\}$;\label{alg:bb10}
\State \hspace{0.1in} {\tt sort\_and\_prune}($Q, \sum C^l, \sum C^u$);\label{alg:bb11}
\State \textbf{return} $(\pi,\alpha)$ that minimizes $\sum C$.\label{alg:bb12}
\EndProcedure
\end{algorithmic}
\end{algorithm}

\vspace{0.05in}\boldstart{Initialization} (line~\ref{alg:bb5}): We initialize the lower and upper bounds for each node according to Lemma~\ref{lemma:bound}. The query accuracy $\prod\alpha$ in Equation~\ref{eq:problem} is within $[{\mathcal{A}^n}, 1]$. For example, for the plan \mathsymbol{XYZ} in Figure~\ref{fig:4bbSearchAlg}, we initialize the lower and upper bounds for each node with $\alpha^l=\mathcal{A}$, $s^l=0$, $r^u=1$ and $\alpha^u=1$, $s^u=1$, $r^l=0$, respectively. The query accuracy $\prod\alpha$ is within $[0.9^3,1]$ initially, where $0.9$ is the query target accuracy $\mathcal{A}$.

\vspace{0.05in}\boldstart{Choosing the next candidate node.} (line~\ref{alg:bb7}): We find the first unvisited tree node $t$ from $\pi$ that is in the front of the queue. In Figure~\ref{fig:4bbSearchAlg}(a), $\pi=XZY$ is in the front of the queue $Q$ according to {\tt sort\_and\_prune()}, which will be explained later. {\tt pop\_unvisited()} yields $\pi=XZY$ and node 3, since node 1 has been visited. Similarly, {\tt pop\_unvisited()} yields $\pi=XZY$ and node 5 in Figure~\ref{fig:4bbSearchAlg}(b). If all the nodes for the head plan in the queue have been visited, we look for the next $\pi\in Q$.

\vspace{0.05in}\boldstart{Tightening cost bounds.} (line~\ref{alg:bb8}$\sim$line~\ref{alg:bb9}): We first call {\tt accuracy\_allocation()} to build an optimal proxy models $\hat{\sigma}^{*}$ with an optimal $\alpha^{*}$ from the root till the current node $t$ at depth $i$. The {\tt update\_node()} function updates $\alpha^l=\alpha^u=\alpha^{*}$ for nodes from the root till $t$. Similarly, $s^l=s^u=s^{*}$, and $r^l=r^u=r^{*}$. This process improves the bounds of $\sum C$ for plans under node $t$ (with untrained $\hat\sigma$s) and in turn tightens the query accuracy $\prod\alpha$ to $[\mathcal{A}^{n-i+1}, \mathcal{A}]$. In Figure~\ref{fig:4bbSearchAlg}(a), for node 3, we call {\tt accuracy\_allocation()} for the sub-query \mathsymbol{XZ} and find the optimal $\alpha_{X}^l=\alpha_{X}^u=0.92$ and $\alpha_{Z}^l=\alpha_{Z}^u=0.98$ for node 1 and node 3, respectively. The {\tt update\_node()} tightens the query accuracy $\prod\alpha$ for the plan \mathsymbol{XZY} from $[0.9^3,1]$ to $[0.9^2,0.9]$, and tightens the lower and upper bounds of $\sum C$ to $[2,8]$.

\vspace{0.05in}\boldstart{Pruning plans.} (line~\ref{alg:bb11}): After the bounds are updated, we sort and prune $\pi\in Q$. The following rules are used to determine the sort order of $\pi$ as well as to prune unnecessary plans. 
\begin{itemize} 
    \item When $[\sum C^l, \sum C^u]$ for two $\pi$'s have overlap, the one with a lower mean cost $\frac{\sum C^l+\sum C^u}{2}$ has a higher priority and is likely to yield more gains. Such a plan should be explored first. In Figure~\ref{fig:4bbSearchAlg}(a), the mean cost for the plan \mathsymbol{XZY} is 5, which is less than that of the plan \mathsymbol{XYZ}. Therefore, the plan \mathsymbol{XZY} has a higher priority than the plan \mathsymbol{XYZ}.
    
    \item When $[\sum C^l, \sum C^u]$ for two $\pi$'s have no overlap, we prune the one with a higher value range from the search tree, since it provides greater cost. In Figure~\ref{fig:4bbSearchAlg}(b), $[\sum C^l, \sum C^u]$ for the plan \mathsymbol{XZY} is lower than that of the plan \mathsymbol{XYZ}, and they have no overlap. Then the plan \mathsymbol{XYZ} is removed from $Q$, i.e., the edge connecting node 2 and node 4 is deleted.
\end{itemize}

The above comparisons are done for each pair of $\pi$'s until $Q$ is fully sorted. The lower bound and upper bound are equal to the exact cost once $\hat\sigma$ is built. Pruned $\pi$'s are removed from $Q$.

\subsection{Improvement Using a Fine-grained Tree}
\label{sec:finegrainedtree}

\begin{figure}[hbt!]
\centering
\includegraphics[width=2.5in]{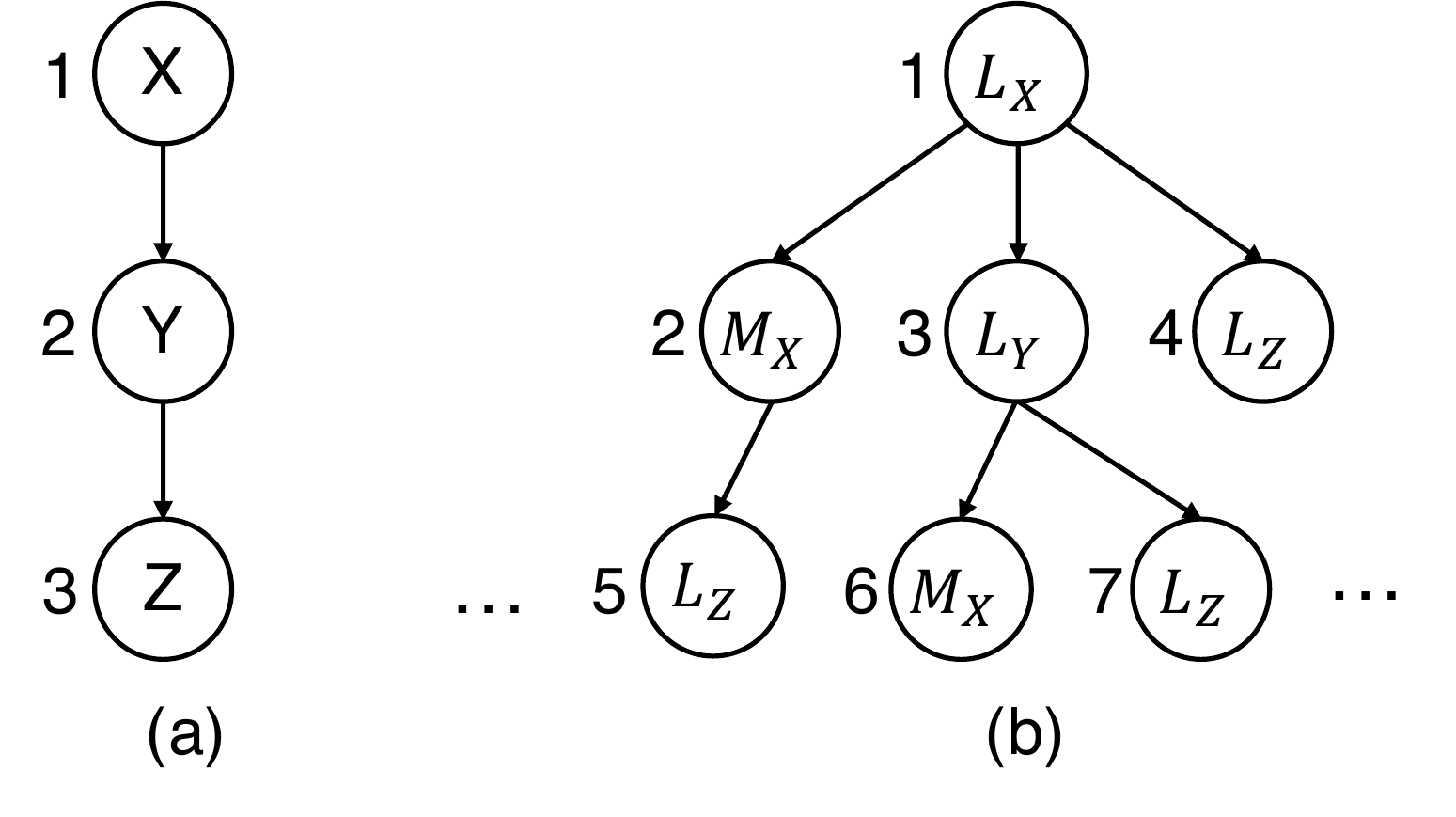}
\caption{(a) A snippet of the search tree in Figure~\ref{fig:4bbSearchAlg}; (b) A fine-grained tree of (a).}
\label{fig:finegrainedtree}
\end{figure}

The branch-and-bound search discussed above involves generating labeled samples ${L}$, followed by training classifiers ${M}$ and deriving $C$ for each node in $\mathbb{H}$. To further speedup the search, we split one node into two: an ${L}$-node to generate labeled samples, and an ${M}$-node to train classifiers ${M}$ and derive ${R}$ and $C$. An ${L}$-node has to be placed before its corresponding ${M}$-node, i.e., labeling happens before training. For instance, the node \mathsymbol{X} in Figure~\ref{fig:finegrainedtree}(a) is split into an $L_{X}$ node to generate the labeled sample for $\hat{\sigma}_{X}$ and an $M_{X}$ node to train the classifier for $\hat{\sigma}_{X}$ in Figure~\ref{fig:finegrainedtree}(b). We call this new tree a \emph{fine-grained} search tree $\mathbb{H}^{+}$.

Compared to the original search tree discussed in the previous section, $\mathbb{H}^{+}$ provides more opportunities to tighten the cost bounds. For example, we can prune the search tree at an ${L}$-node without executing its corresponding ${M}$-node. The search algorithm is similar to Algorithm~\ref{alg:B-B-pruning}, except a new {\tt update\_node()} function.  Its update scheme now depends on the type of node $t$, discussed below.

\vspace{0.05in}
\noindent{${L}$-node.} We update the lower and upper bounds of selectivity $s$ because we generate labeled samples and compute $s$ at $L$-node. For an $L$-node $t$, a proxy model $\hat{\sigma}$ is called \emph{available for $t$} if its corresponding $M$-node is an ancestor of $t$; otherwise, $\hat{\sigma}$ is called \emph{unavailable for $t$}. We compute lower and upper bounds of ${s_t}$ by applying all available prefix $\hat\sigma$ and $\sigma$ on the raw input to obtain a labeled sample $L^{*}_t$, and its selectivity is denoted as $s^{*}_t$. In Figure~\ref{fig:finegrainedtree}(b), $\hat{\sigma}_X$ is available for node 5 because we build $\hat{\sigma}_X$ at node 2, which is an ancestor of node 5, while it is unavailable for node 3 because $M_X$ is not an ancestor of node 3. The labeled sample $L^{*}_{Y}$ for node 3 is labeled by $\sigma_{Y}$ after $\sigma_{X}$ on the raw input without applying $\hat\sigma_{X}$. Let the selectivity on $L^{*}_{Y}$ be $s^{*}_{Y}$. We compute $C_t^l$ and $C_t^u$ as follows:
\begin{itemize}
\item A lower bound $C_t^l$ can be computed when its unavailable proxy models have $\alpha^l=\mathcal{A}$ and discard records that satisfy ${\sigma_t}$ from $L_t^{*}$. In this case, the selectivity $s$ becomes ${(s_t^{*} - (1-\mathcal{A})^k)}$ ${/\mathcal{A}^k}$, where $k$ is the number of unavailable prefix proxy models. This selectivity is used to estimate ${C_t^l}$ using Expression~\ref{eq:lowerBound}. For node 3 in Figure~\ref{fig:finegrainedtree}(b), we compute $C_Y^l$ using $s_Y^l=(s_Y^{*}-(1-\mathcal{A}))/\mathcal{A}$ when the unavailable $\hat{\sigma}_X$ with $\alpha=\mathcal{A}$ discards records satisfying $\sigma_Y$ from $L_Y^{*}$.
\item An upper bound $C_t^u$ can be computed when unavailable proxy models do not discard any records in ${L^{*}_t}$ (i.e., $\alpha=1.0$). Its selectivity is $s_t^{*}$ in this case. We compute $C_t^u$ using $s_t^u={s^{*}_t}$ in Expression~\ref{eq:upperBound}. In Figure~\ref{fig:finegrainedtree}(b), at node 3, when $\hat{\sigma}_{X}$ is unavailable and we use $\alpha=1.0$, the selectivity $s_Y^u=s^{*}_{Y}$ is used to estimate $C_{Y}^u$.
\end{itemize}

\vspace{0.05in}
\noindent\emph{$M$-node.} As in Section~\ref{sec:branchandbound}, we call Algorithm~\ref{alg:accuracy-allocation} to compute ${\alpha^{*}}$, train $\hat\sigma$, and estimate ${C}$. We also update the bounds for all its ancestor nodes. In Figure~\ref{fig:finegrainedtree}(b), after we train $\hat{\sigma}_{X}$ for node 5, we update the selectivity of node 3 by applying $\hat{\sigma}_{X}^l$ on its labeled sample $L'_{Y}$.

\vspace{0.05in}
The above search on the fine-grained tree is efficient, as illustrated in our experiments. For a query on the Twitter dataset, the search algorithm prunes 37\% of the nodes on the original search tree, and 85\% of the nodes on the fine-grained tree.

\section{experiments}
\label{sec:experiments}

We have conducted a thorough evaluation of {\name} and compared it with state-of-the-art solutions.

\subsection{Setup} 
\label{sec:setup}

\vspace{0.05in}\noindent\textbf{Datasets.} We used three datasets with text, images, and videos. 

\vspace{0.05in}
\noindent\emph{Twitter text dataset.} It contained 2M tweets from January 2017 to September 2017 in the United States randomly sampled using the Twitter sampled stream API~\cite{TwitterAPI}. Each tweet was a string with a maximum of 140 characters. This dataset supported text analysis and retrieval by utilizing various NLP modules such as entity recognition, sentiment analysis, and part-of-speech (PoS) tagger. 

\vspace{0.05in}
\noindent\emph{COCO image dataset.} COCO~\cite{lin2014microsoft} was a public dataset collected online. It contained 123K images and 80 object classes such as ``person'', ``bicycle'', and ``dog''. Each image was labeled with multiple objects for their class labels and bounding box positions. The dataset was used for retrieving images that contained one or more object classes specified in user queries. 

\vspace{0.05in}
\noindent\emph{UCF101 video dataset.} The UCF101 activity recognition dataset~\cite{soomro2012ucf101} contained 13K videos collected from YouTube. Each video was labeled with one of 101 action categories such as ``applying lipstick'' and ``baby crawling''. It supported video retrieval using labels generated by object detection and action recognition models. 

\begin{figure}[hbt!]
\subfloat[$q_1$ on the Twitter dataset.]{\includegraphics[width=3.4in]{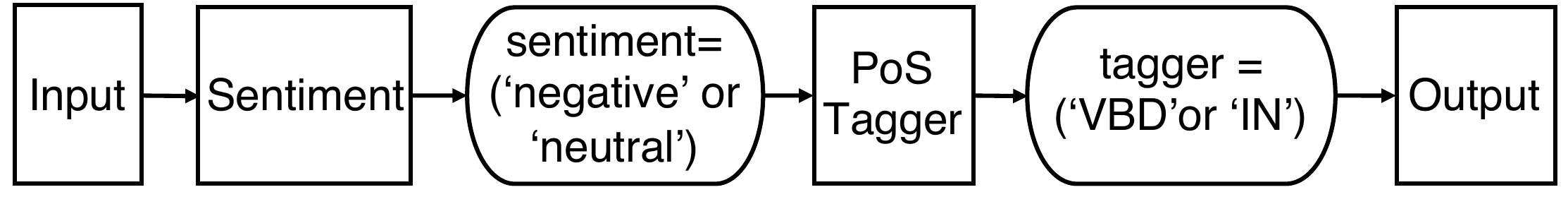}}
\\
\subfloat[$q_2$ on the COCO dataset.]{\includegraphics[width=3.4in]{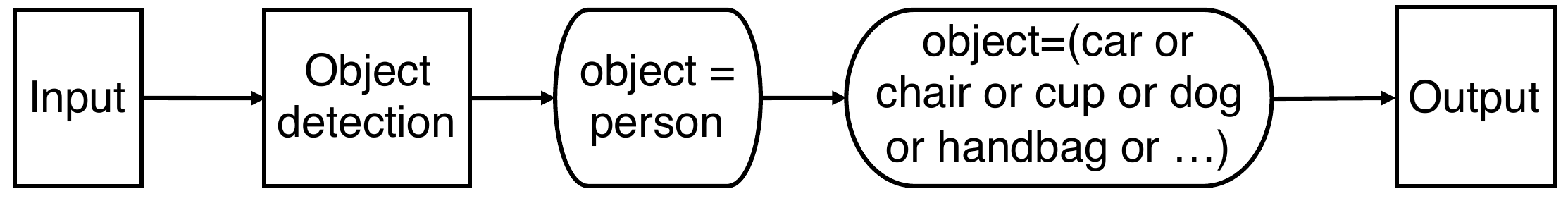}}
\caption{Sample ML workflows.}
\label{fig:6exp1workflow}
\end{figure}

\vspace{0.05in}
\noindent\textbf{Workloads.} To our best knowledge, there is no off-the-shelf benchmark for ML inference with comprehensive ML operators and predicates. To solve the problem, we generated 10 queries for each dataset in the experiments. Table~\ref{tab:workflows} illustrates some of them, and Figure~\ref{fig:6exp1workflow} shows some sample workflows. The workloads retrieved texts, images, and videos that matched given query predicates, which were conjunctions of multiple clauses with different selectivity values. Each predicate clause was an equality condition on an ML-generated label column. We refer the readers to a full list of the queries as well as snapshots of the datasets in~\cite{CorrelativeCascades}. Each query also specified a target query accuracy ${\mathcal{A}}$, indicating how much accuracy loss the user was willing to pay relatively to the original query. 

\begin{table}[hbt!]
\footnotesize
\centering
\begin{tabular}{@{\hspace{0.01cm}}l@{}|@{\hspace{0.05cm}}l@{}|m{4.5cm}|@{\hspace{0.05cm}}l@{}|@{\hspace{0.05cm}}l@{}}
Dataset & Q\# & Query semantics & Selectivity & Correlation\\\hline
\hline
\multirow{2}{*}{\shortstack[l]{Twitter}} & q${_1}$ & Sentiment('negative' or 'neutral') \& PoS Tagger('VBD' or 'WRB' or 'IN')& 0.49& 0.55\\
\cline{2-5}
 & q${_2}$ & Sentiment('negative' or 'neutral') \& PoS Tagger('PRP') & 0.35 & 0.41\\
\hline
\multirow{2}{*}{\shortstack[l]{COCO}} & q${_2}$ & Object detection (person) \& (car or chair or cup or dog or handbag or $\dots$) & 0.20 & 0.98\\
\cline{2-5}
 & q${_6}$ & Object detection (person) \& (car or chair or cup or tv or bed or $\dots$) & 0.13 & 0.99\\
\hline
\multirow{2}{*}{\shortstack[l]{UCF101}} & q${_2}$ & Activity Recognition (archery or balance beam or biking or $\dots$) \& Object detection (chair or sports ball or bird or $\dots$)& 0.17 & 1.00\\
\cline{2-5}
 & q${_9}$ & Object detection (chair or sports ball or cup or $\dots$) \& Activity Recognition (archery or balance beam or basketball dunk or $\dots$) & 0.22 & 0.82\\
\hline
\end{tabular}
\caption{Some of ML queries used in the experiments. }
\label{tab:workflows}
\end{table}

\vspace{0.05in}
\noindent\textbf{Metrics.} We measured (1) the end-to-end total processing time that included the query optimization, training of necessary models, and processing the query given an optimized plan; (2) the accuracy of our query processing relatively to the original ML inference queries; (3) the query execution cost (milliseconds per record); and (4) the decomposition of the optimization costs (minutes). 

\vspace{0.05in}

\noindent\textbf{CORE.} We implemented a query execution engine and the {\name} optimizer in Python that enabled ML inference queries on various unstructured texts, images, and videos.  We also implemented several ML UDFs using the Stanford NLP~\cite{manning2014stanford} and spaCy packages for text analysis, YOLOv3~\cite{yolov3} for object detection in images, and an activity recognition model~\cite{activityRecognition} for recognizing activities in videos. 

\vspace{0.05in}
\noindent\textbf{Baselines.} We compared {\name} against the following baseline approaches. (i) \textsf{ORIG} was a baseline that ran the original query as it is. (ii) \textsf{NS} was a baseline based on NoScope~\cite{kang2017noscope}.  It trained a single light-weight model and inserted it early in a plan to quickly filter input records that did not match the query predicate so that the entire query could be accelerated. (iii) \textsf{PP} (short for Probabilistic Predicates~\cite{lu2018accelerating}) built a light-weight filter for each predicate offline and injected them early in a plan with an independence assumption of predicates, given an ad-hoc query.

The experiments were run on a c5.4xlarge AWS instance with 280GB SSD storage, 16 vCPUs, and 32GB memory, running a Ubuntu Linux 16.04.

\subsection{Effect of Predicate Correlation}
\label{sec:6exp62}
To understand the effect of correlation of UDFs in a query, we leveraged the Twitter dataset and 20 test queries with two or three predicates. The queries were divided by their correlation score ${\hat{\kappa}^2}$ at a cutoff score of 0.2. As a result, each query was classified as \emph{weakly} or \emph{strongly} correlated among the predicates according to $\hat{\kappa}^2$. Table~\ref{tab:queryCorrelationScore} shows the correlation scores of these queries. We collected the execution costs of these weakly and strongly correlated queries with a query accuracy $\mathcal{A}=90\%$. We ran these queries using {\sf ORIG}, {\sf NS}, {\sf PP}, and {\name} to generate optimal plans, and tested the execution cost of an optimal plan by executing the plan on a sample of data.

\begin{figure}[hbt!]
\centering
\includegraphics[width=3.2in]{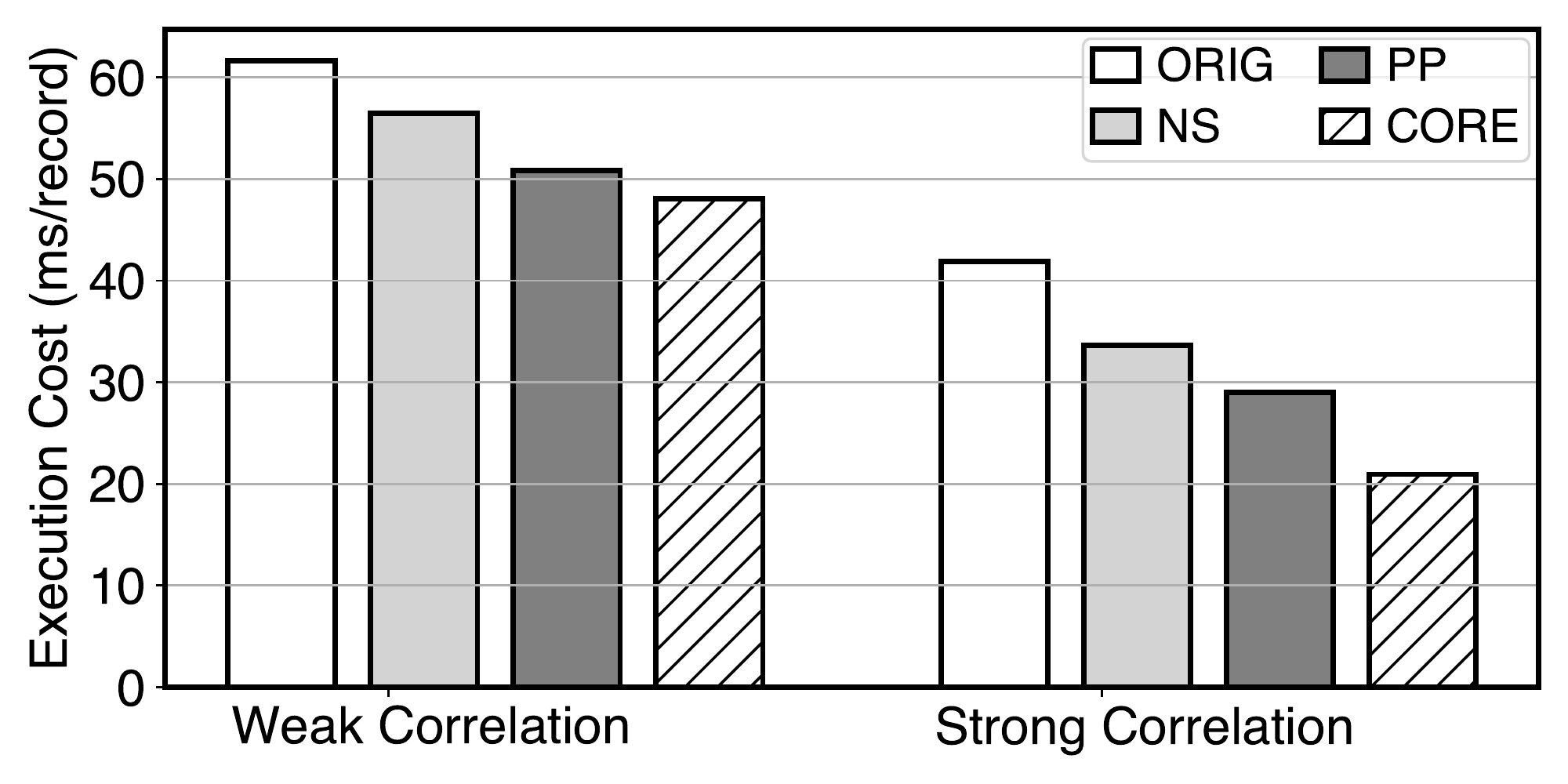}
\caption{Average execution costs over two sets of queries $q_1\sim q_{10}$ and $q'_1\sim q'_{10}$ with strong and weak correlations, respectively.}
\label{fig:correlation}
\end{figure}

Figure~\ref{fig:correlation} shows the execution costs. From Figure~\ref{fig:correlation}, we can see that (i) {\sf NS}, {\sf PP}, and {\name} reduced the execution cost compared to {\sf ORIG}, and (ii) compared to {\sf PP}, {\name} reduced the execution cost more on strongly correlated queries than weakly correlated queries. In general, {\sf NS} improved over {\sf ORIG} using cheap filters to quickly discard irrelevant inputs, and {\sf PP} further boosted the performance by decomposing the filters according to the predicate clauses. Note that there was still room for improvements for queries with more correlations and {\name} filled this gap as expected.

\begin{table}[hbt!]
\small
\centering
\begin{tabular}{m{0.7cm}|m{0.35cm}|m{0.35cm}|m{0.35cm}|m{0.35cm}|m{0.4cm}|m{0.4cm}|m{0.4cm}|m{0.4cm}|m{0.4cm}|m{0.4cm}}
\hline
 & ${q'_1}$ & ${q'_2}$ & ${q'_3}$ & ${q'_4}$ & ${q'_5}$ & ${q'_6}$ & ${q'_7}$ & ${q'_8}$ & ${q'_9}$ & ${q'_{10}}$ \\
\hline
Weak & 0.15 & 0.15 & 0.15 & 0.15 & 0.16 & 0.16 & 0.16 & 0.16 & 0.16 & 0.16\\ 
\hline
\end{tabular}
\begin{tabular}{m{0.7cm}|m{0.35cm}|m{0.35cm}|m{0.35cm}|m{0.35cm}|m{0.4cm}|m{0.4cm}|m{0.4cm}|m{0.4cm}|m{0.4cm}|m{0.4cm}}
\hline
& ${q_1}$ & ${q_2}$ & ${q_3}$ & ${q_4}$ & ${q_5}$ & ${q_6}$ & ${q_7}$ & ${q_8}$ & ${q_9}$ & ${q_{10}}$ \\
\hline
Strong & 0.55 & 0.41 & 0.55 & 0.42 & 0.41 & 1.00 & 0.80 & 0.96 & 0.80 & 0.93\\
\hline
\end{tabular}
\caption{The correlation scores for ten strongly correlated queries $q_1\sim q_{10}$ (marked as ``Strong'') and ten weakly correlated queries $q'_{1}\sim q'_{10}$ (marked as ``Weak'') on the Twitter dataset.}
\label{tab:queryCorrelationScore}
\end{table}

\subsection{Time Reduction of {\name}}
\label{sec:6exp63}
To study the performance improvements of {\name} over existing solutions, we tested the total times of strongly correlated queries with $\mathcal{A}=90\%$ on the three datasets. For query optimization to generate an optimal query plan, we used 0.34\% of the input data on the Twitter dataset, 0.84\% of the input data on the COCO dataset, and 14.86\% of the input on the UCF101 dataset (due to its smaller size). After generating the optimal plan, we ran it on the rest of the input. The total time included the optimization time and the time of processing all the records.  We used the same setting for {\sf NS} and {\sf PP}, which built proxy models online.

\begin{figure}[hbt!]
\subfloat[Total time (Twitter).\label{fig:6exp2exeCostTwitter}]{\includegraphics[width=1.6in]{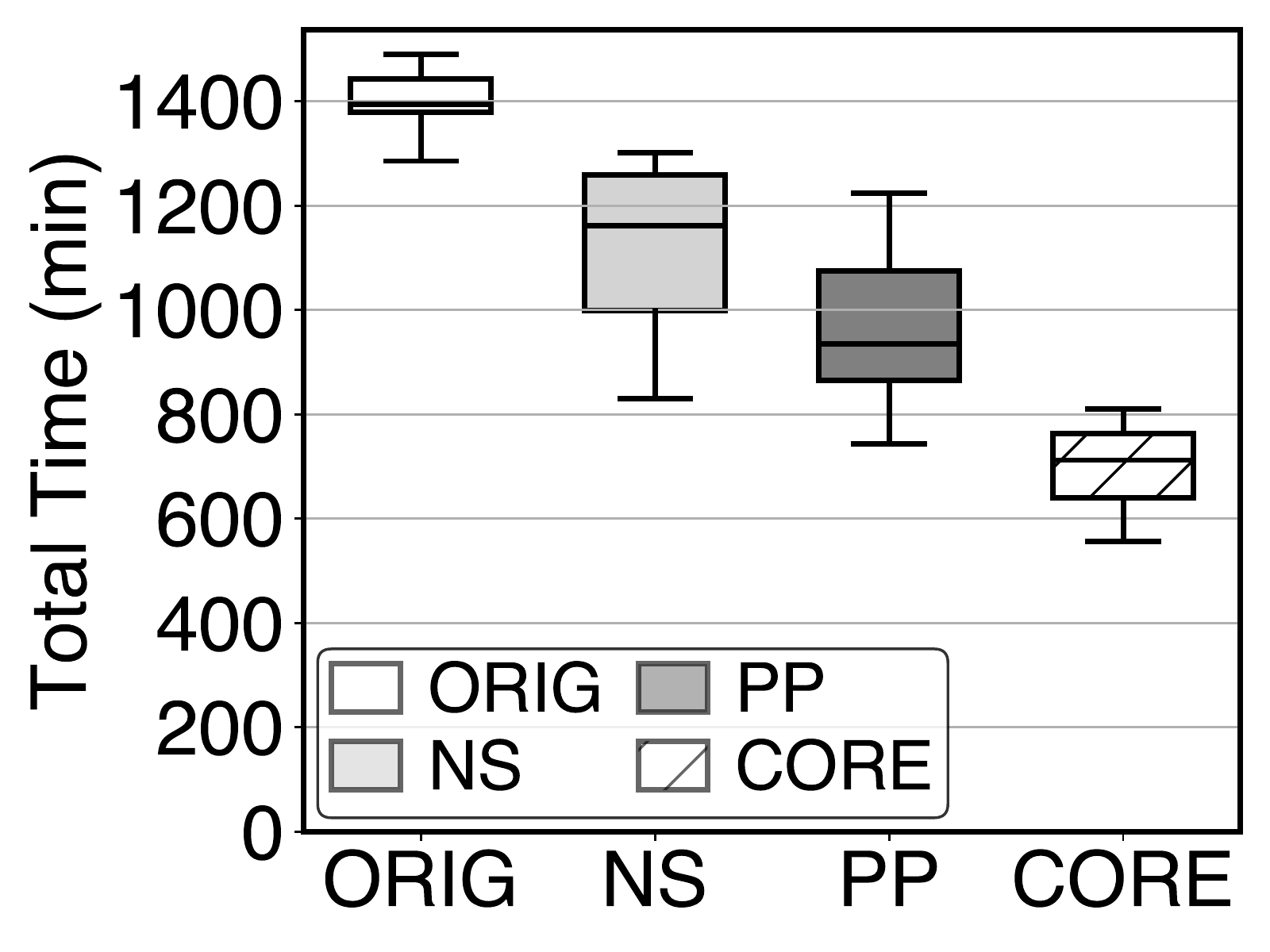}}
\subfloat[Time reduction (Twitter).\label{fig:6exp2reductTwitter}]{\includegraphics[width=1.6in]{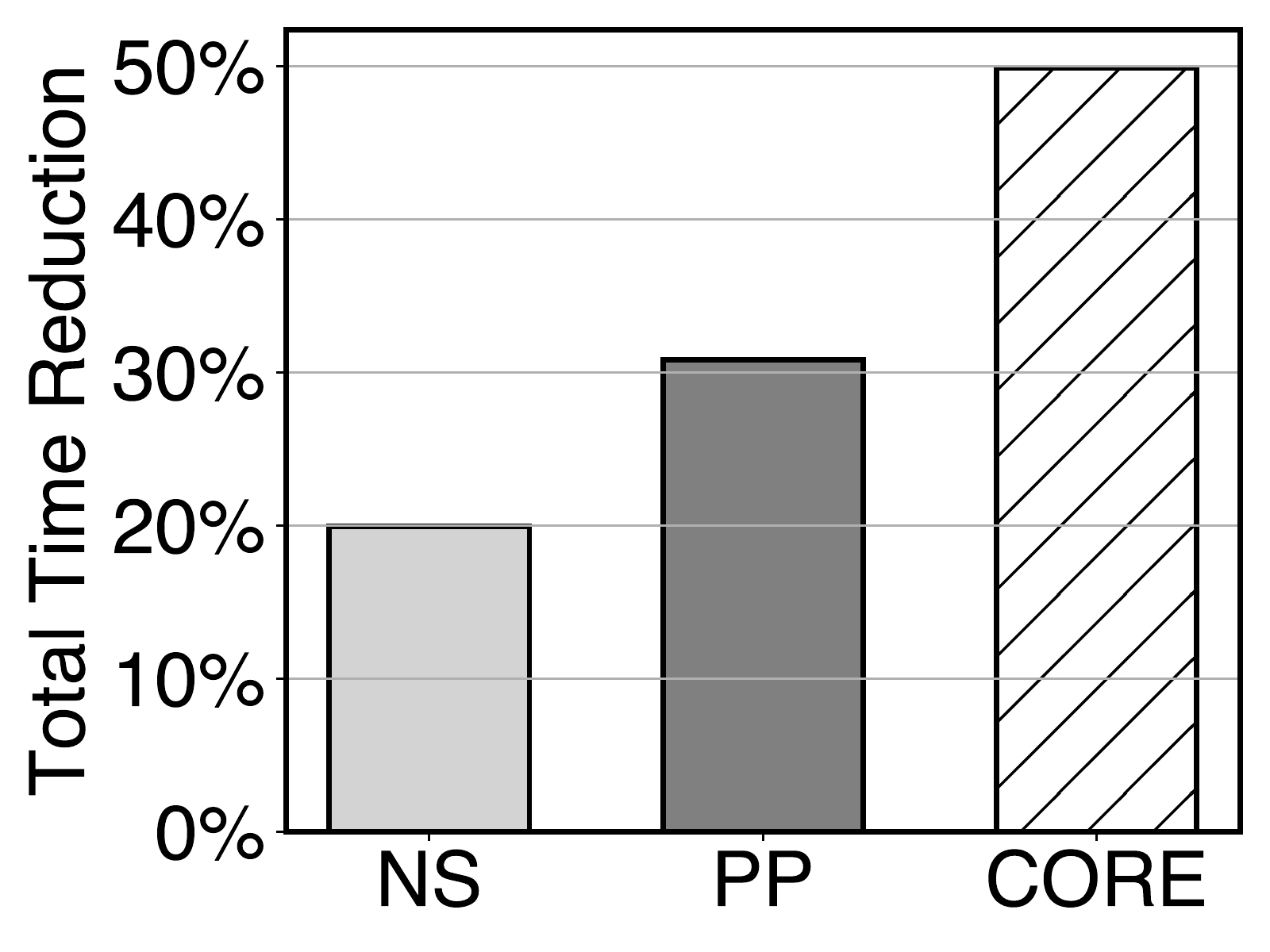}}
\\
\subfloat[Total time (COCO).\label{fig:6exp2exeCostCoco}]{\includegraphics[width=1.6in]{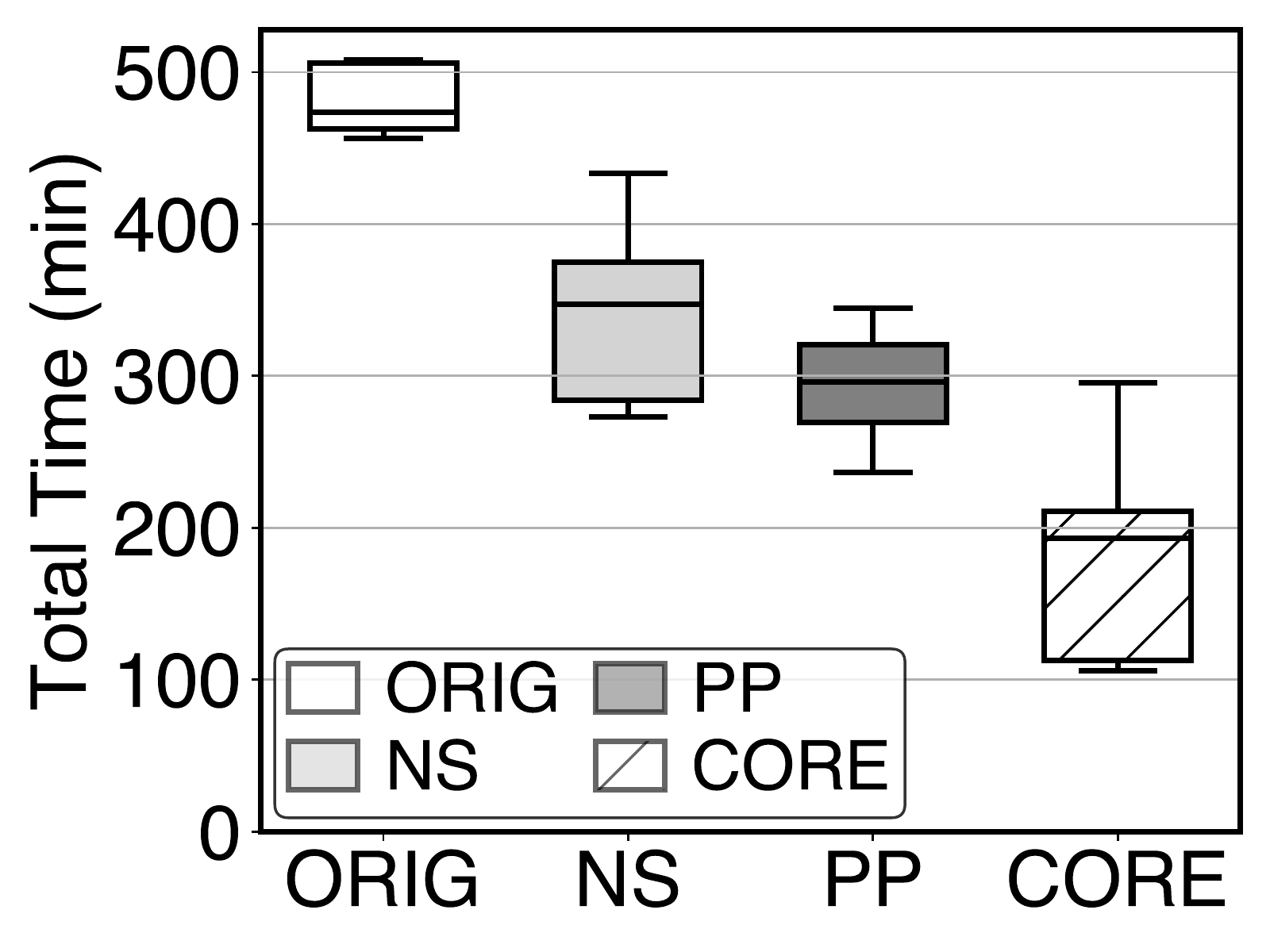}}
\subfloat[Time reduction (COCO).\label{fig:6exp2reductCoco}]{\includegraphics[width=1.6in]{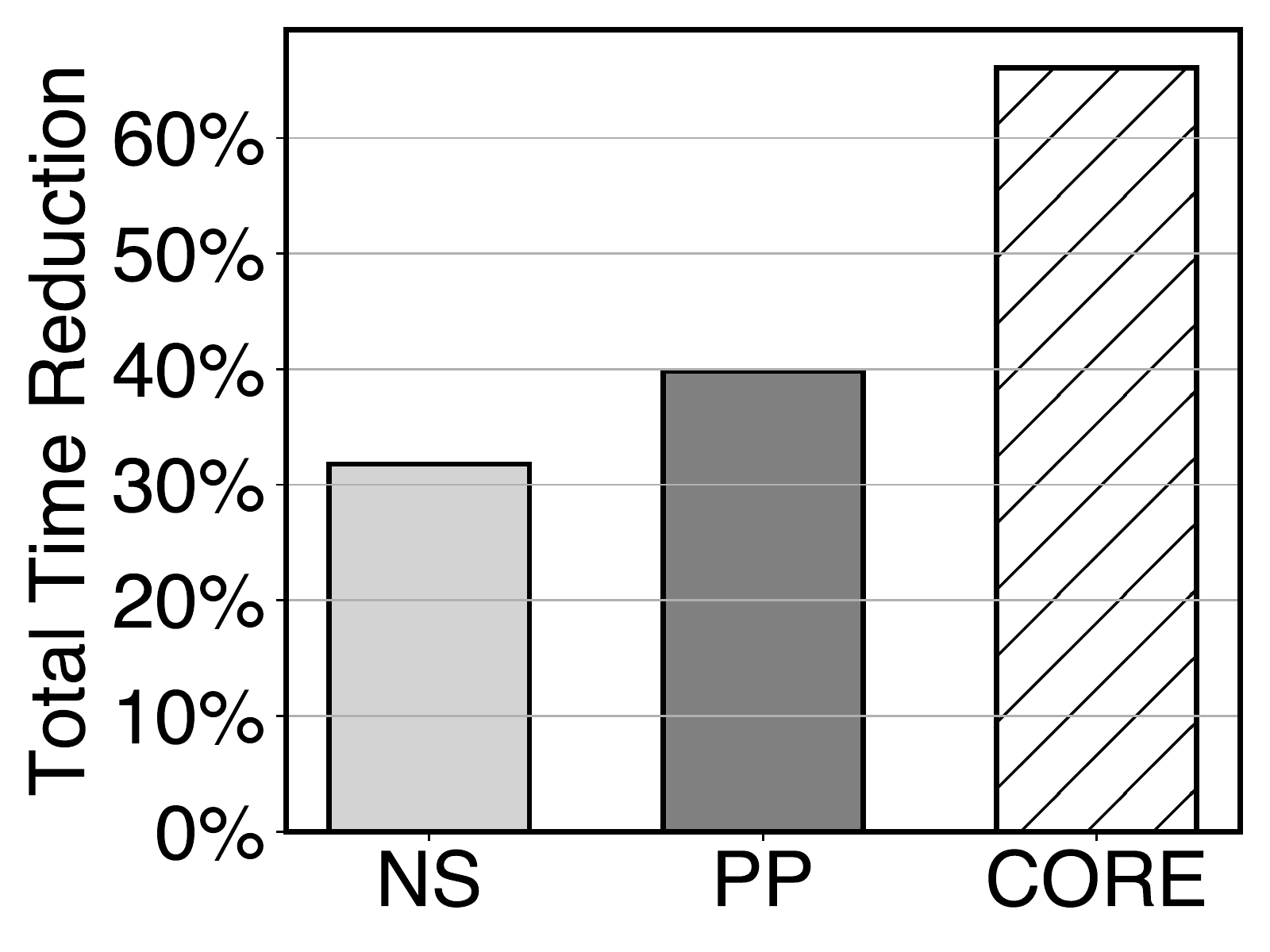}}
\\
\subfloat[Total time (UCF101).\label{fig:6exp2exeCostUcf101}]{\includegraphics[width=1.6in]{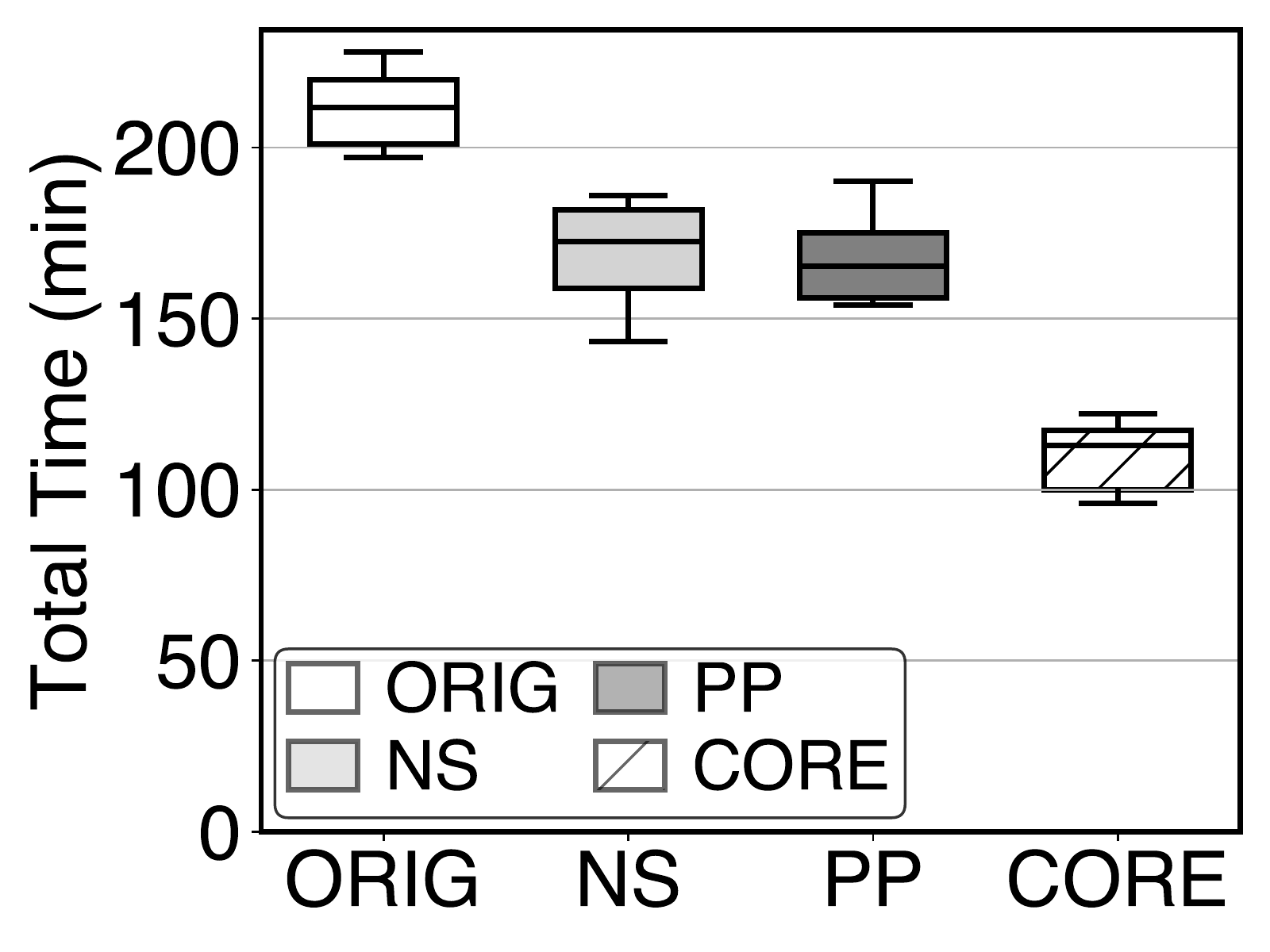}}
\subfloat[Time reduction (UCF101).\label{fig:6exp2reductUcf101}]{\includegraphics[width=1.6in]{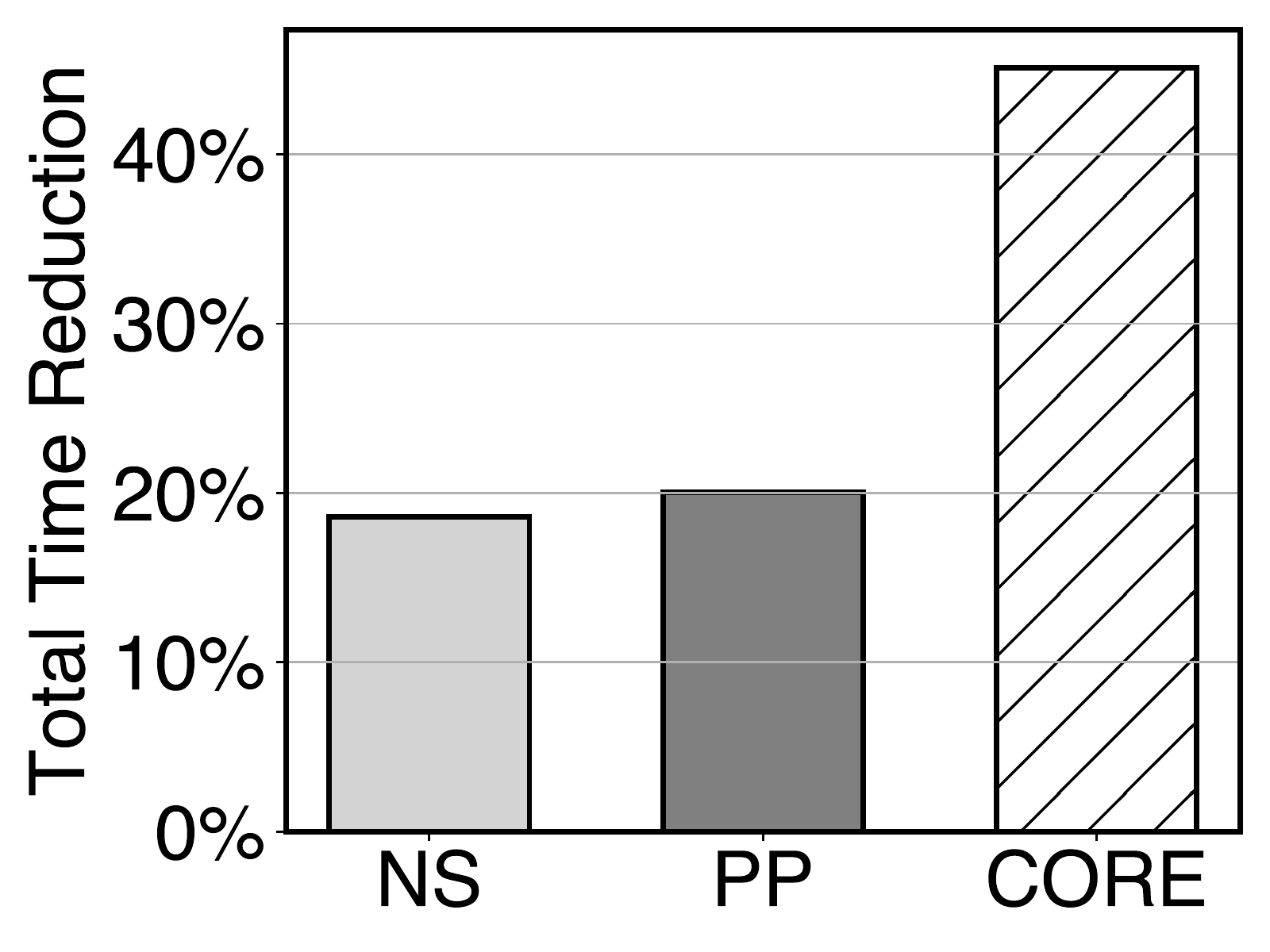}}
\caption{The total time over ten queries for each dataset using {\name} and baseline approaches. ${\mathcal{A}=90\%}$. For (a), (c), and (e), we show the $1^{st}$ and $99^{th}$ percentiles on the bars and $1^{st}$ quartile, median, and $3^{rd}$ quartile on the boxes. For (b), (d) and (f), we present the average total time reductions relative to {\sf ORIG}.}
\label{fig:6exp2exeCost3sets}
\end{figure}

\begin{figure*}[hbt!]
\includegraphics[width=\linewidth]{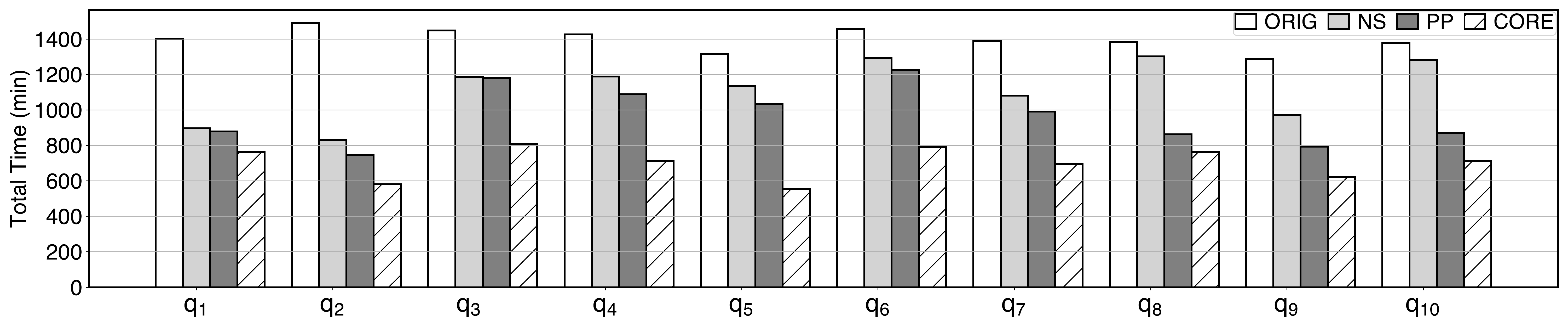}
\caption{The total time of each query in the Twitter dataset using {\sf ORIG}, {\sf NS}, {\sf PP}, and {\name}.}
\label{fig:exp2exeCostTwitter}
\end{figure*}

\cref{fig:6exp2exeCostTwitter,fig:6exp2exeCostCoco,fig:6exp2exeCostUcf101} show the total times of ten queries in each dataset, and~\cref{fig:6exp2reductTwitter,fig:6exp2reductCoco,fig:6exp2reductUcf101} show the average total-time reductions for the ten queries using {\sf NS}, {\sf PP}, and {\name} compared to {\sf ORIG}. We also presented the total time of each individual query in the Twitter dataset in Figure~\ref{fig:exp2exeCostTwitter}. These results show that {\name} had a better performance than the baseline approaches in general. Specifically, {\name} achieved up to a 61\% reduction on the Twitter dataset compared to {\sf ORIG}. For {\sf NS} and {\sf PP}, the reductions were about 44\% and 50\%, respectively. We observe similar reductions on other datasets as well. For example, on the COCO dataset, {\name} had a reduction of  up to 73\% compared to {\sf ORIG}, while {\sf NS} and {\sf PP} achieved a reduction of 35\% and 44\%, respectively. In addition, the average/variance correlation scores for strongly correlated queries were 0.68/0.6 on the Twitter dataset, 0.99/0.01 on the COCO dataset, and 0.94/0.01 on the UCF101 dataset. As discussed in Section~\ref{sec:correlation}, {\name} achieved more gains over {\sf PP} when the queries had predicates with a stronger correlation. 

\subsection{Optimization Cost of {\name}} 

To better understand the detailed optimization cost of {\name}, we collected the time to generate labeled samples, the time to train classifiers, and the time of search frameworks for each query. The optimizer {\name} used multiple threads to label training samples. Each ML model processing unstructured texts used ten threads in parallel. The YOLOv3 model and the image feature model used two processes in parallel, and the activity recognition model used six processes in parallel. During the phase of building proxy models, the size of labeled sample $L$ was empirically set to $1,000$. The training set, testing set, and validation set were split in a 6:2:2 ratio. We also re-sampled the training data to ensure a label balance. We used scikit-learn to train a linear SVM classifier ${M}$ on the labeled sample for text analytic queries, and used keras to train a shallow NN classifier for analytic queries on images and videos. During training, we leveraged a grid-search on the F1-score to decide the best set of hyper-parameters and a cross-validation to train a classifier using the set of hyper-parameters. Additionally, a hill climbing search framework was adopted to find an optimal solution. 

\begin{table}[hbt!]
\footnotesize
\centering
\begin{tabular}{@{\hspace{0.05cm}}c@{\hspace{0.05cm}}|@{\hspace{0.05cm}}c@{\hspace{0.05cm}}|@{\hspace{0.05cm}}c@{\hspace{0.05cm}}|@{\hspace{0.05cm}}c@{\hspace{0.05cm}}|@{\hspace{0.05cm}}c@{\hspace{0.05cm}}|@{\hspace{0.05cm}}c@{\hspace{0.05cm}}|@{\hspace{0.05cm}}c@{\hspace{0.05cm}}|@{\hspace{0.05cm}}c@{\hspace{0.05cm}}|@{\hspace{0.05cm}}c@{\hspace{0.05cm}}|@{\hspace{0.05cm}}c@{\hspace{0.05cm}}}
\hline
&&&Labeling&Training&Searching&QO&QO&Total&Total Time\\
Dataset&ID&\#preds&Time&Time&Time&Time&Time&Time&Reduction\\
&&&(min)&(min)&(min)&(min)&pct.&(min)&(\%)\\
\hline
Twitter & q$_1$ & 2 & 0.93 & 0.10 & 0.17 & 1.20 & 0.16\% & 763 & 45.56\\ 
Twitter & q$_2$ & 2 & 1.22 & 0.09 & 0.14 & 1.46 & 0.25\% & 581 & 60.99 \\
Twitter & q$_8$ & 3 & 1.53 & 0.75 & 3.28 & 5.58 & 0.73\% & 764 & 44.77 \\
Twitter & q$_{10}$ & 3 & 1.76 & 0.75 & 2.93 & 5.47 & 0.77\% & 712 & 48.26 \\
Twitter & Avg. & 2.5 & 1.84 & 0.44 & 2.61 & 4.91 & 0.70\% & 700 & 49.87\\
\hline
COCO & Avg. & 2 & 6.00 & 2.06 & 0.24 & 8.30 & 5.67\% & 173 &66.07\\
\hline
UCF101 & Avg. & 2 & 23.40 & 0.08 & 0.20 & 23.68 & 21.80\% & 110 &49.49\\
\hline
\end{tabular}
\caption{Optimization costs and the total processing time for ten queries over each dataset using {\name} with $\mathcal{A}=90\%$. The ``labeling time'' is the time to generate labeled samples. The ``training time'' is the time to train classifiers. The ``searching time'' is the elapsed time for the search framework. The ``QO time'' is the total time of the labeling, training and searching times. The ``QO Time pct.'' is the percentage of the QO time over the total processing time. Total Time Reduction = ({\sf ORIG}-{\name})/{\sf ORIG}.}
\label{tab:6detailoverhead}
\end{table}

Table~\ref{tab:6detailoverhead} shows the results of the ten queries over each dataset, including the time reduction compared to {\sf ORIG}. On the Twitter dataset, the optimization time was $0.70\%$ of the total time, and the total time reduction was $49.87\%$ on the average. On the COCO dataset, the optimization time was $5.67\%$ of the total time, and the total time reduction was $66.07\%$ on the average. UCF101 was relatively smaller, and $14.86\%$ of the data was used for optimization.  The optimization time was $21.80\%$ of the total time, and the total time reduction was $49.49\%$ on the average. Overall, the query optimization cost of {\name} was a small portion of the total processing time, and it achieved significant performance improvement compared to {\sf ORIG}. When the dataset was small (e.g., the UDF101 dataset) or queries had many ML operators and predicates (e.g., $q_8$ and $q_{10}$ on the Twitter dataset), the query optimization costs were larger.

\subsection{Effectiveness of {\name} Components}
{\name} searched an optimal query plan in both the accuracy space $\mathbb{A}$ and the order space $\mathbb{H}$.  We evaluated the effectiveness of different components in {\name} using two variants, namely {\name}-a and {\name}-h. {\name}-a represented the setting with the reordering step disabled during optimization and constrained the search space to solely $\mathbb{A}$ (Section~\ref{sec:allocateaccuracy}). It used the input-query order and derived an optimal set of accuracy values in $\mathbb{A}$ using Algorithm~\ref{alg:accuracy-allocation}. {\name}-h applied Algorithm~\ref{alg:accuracy-allocation}, and exhaustively searched an optimal order in $\mathbb{H}$ instead of performing the brand-and-bound pruning in Algorithm~\ref{alg:B-B-pruning}.

\begin{figure}[hbt!]
\subfloat[Execution cost (Twitter).\label{fig:exp3exeCostTwitter}]{\includegraphics[width=1.6in]{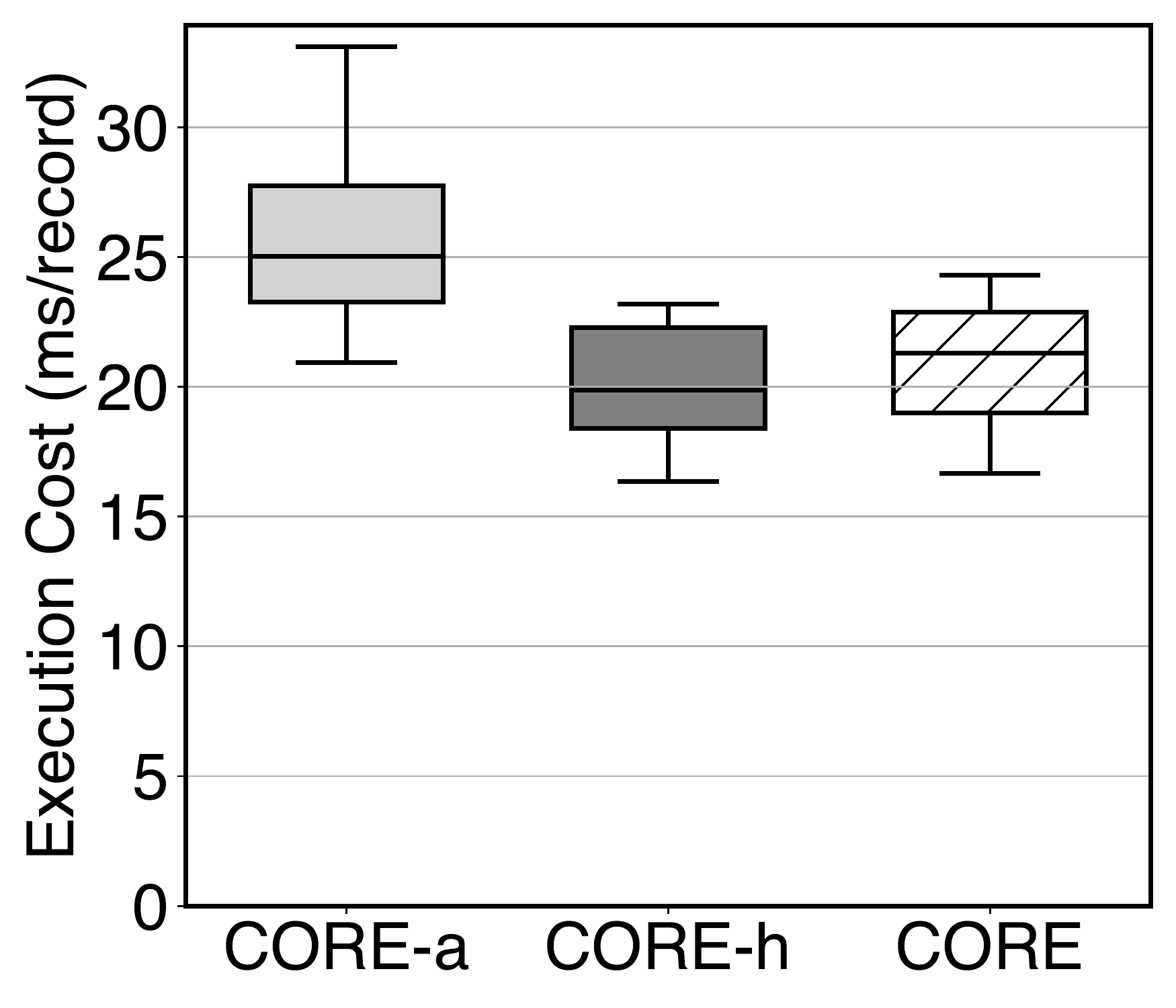}}
\subfloat[Optimization cost (Twitter).\label{fig:exp3optCostTwitter}]{\includegraphics[width=1.6in]{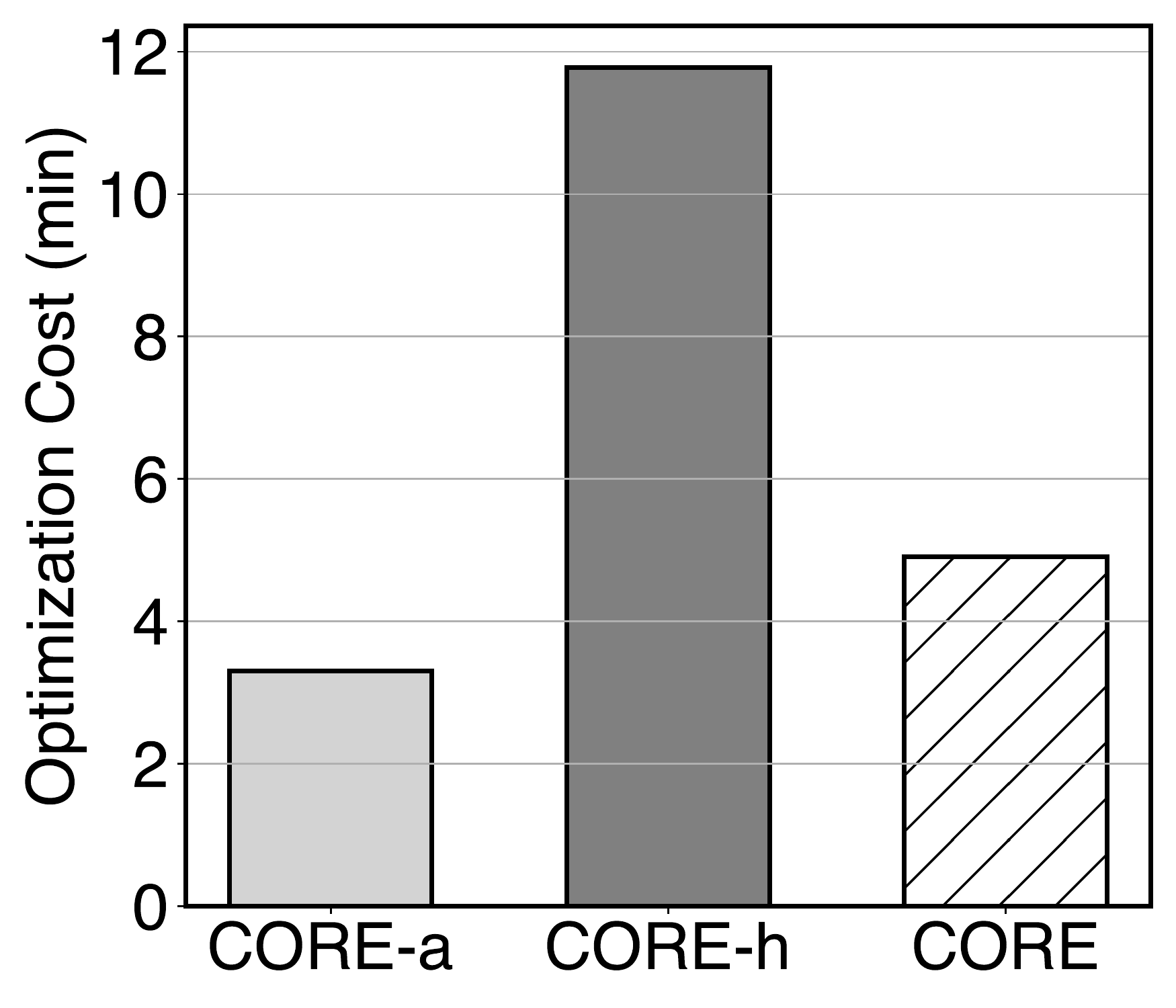}}
\\
\subfloat[Execution cost (COCO).\label{fig:exp3exeCostCoco}]{\includegraphics[width=1.6in]{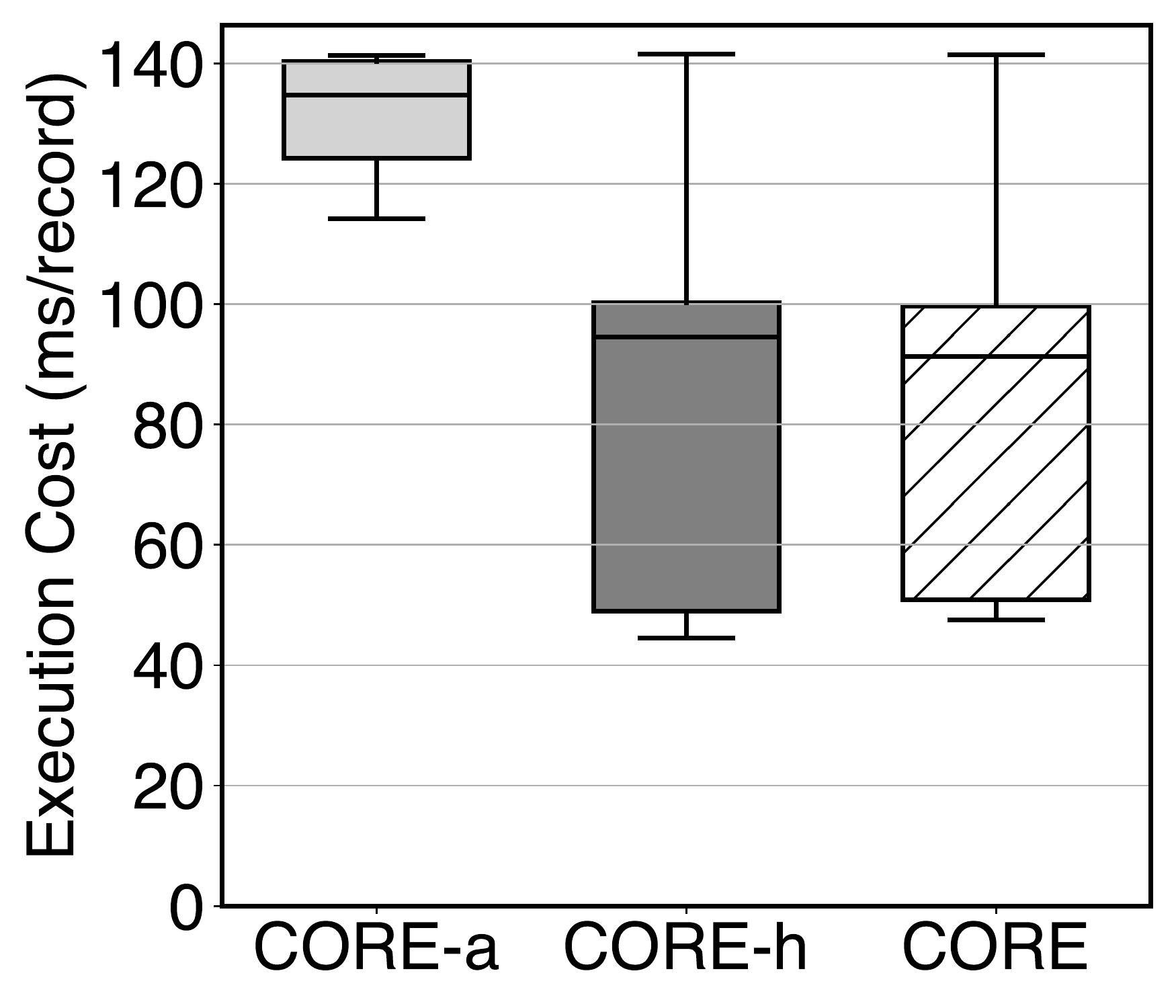}}
\subfloat[Optimization cost (COCO).\label{fig:exp3optCostCoco}]{\includegraphics[width=1.6in]{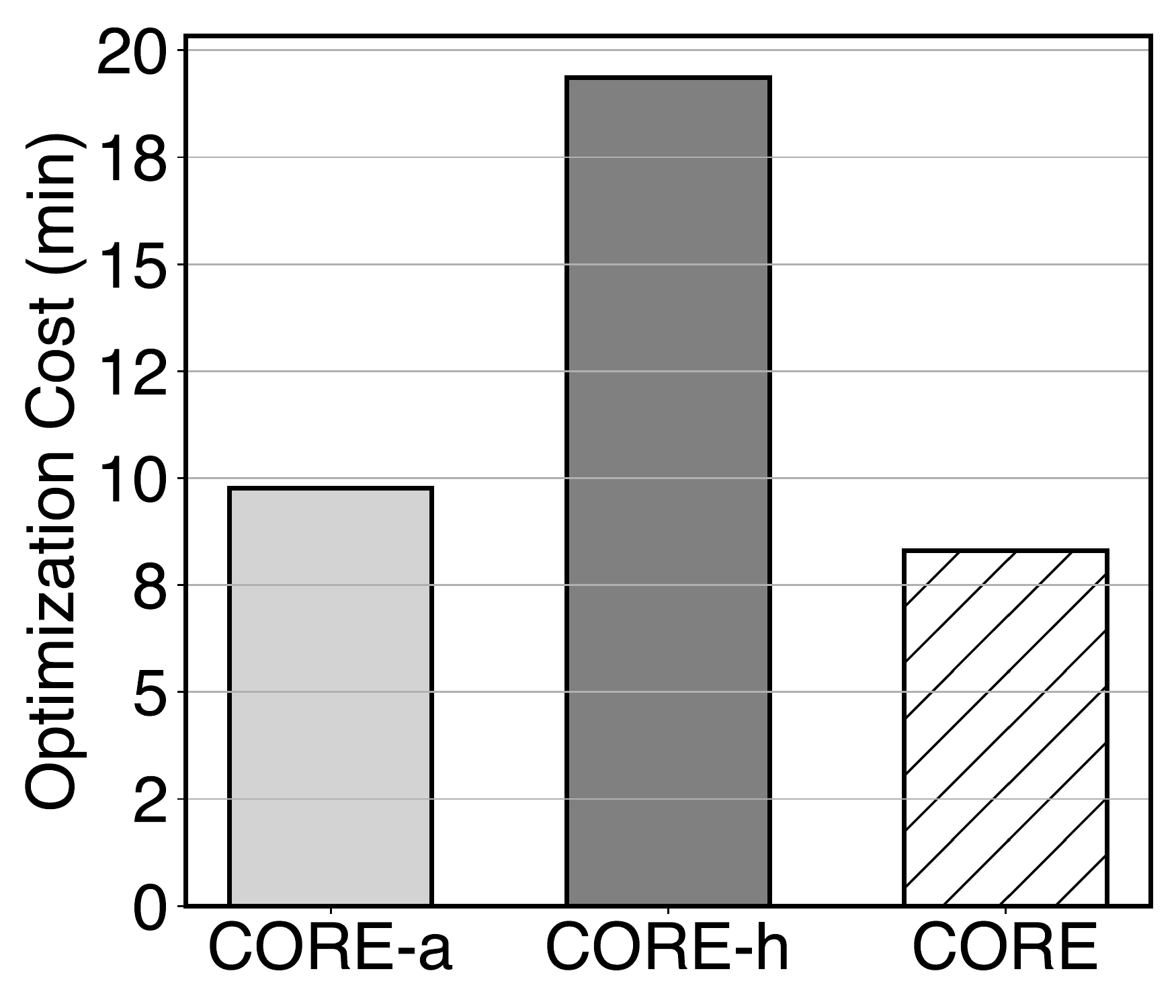}}
\\
\subfloat[Execution cost (UCF101).\label{fig:exp3exeCostUcf101}]{\includegraphics[width=1.6in]{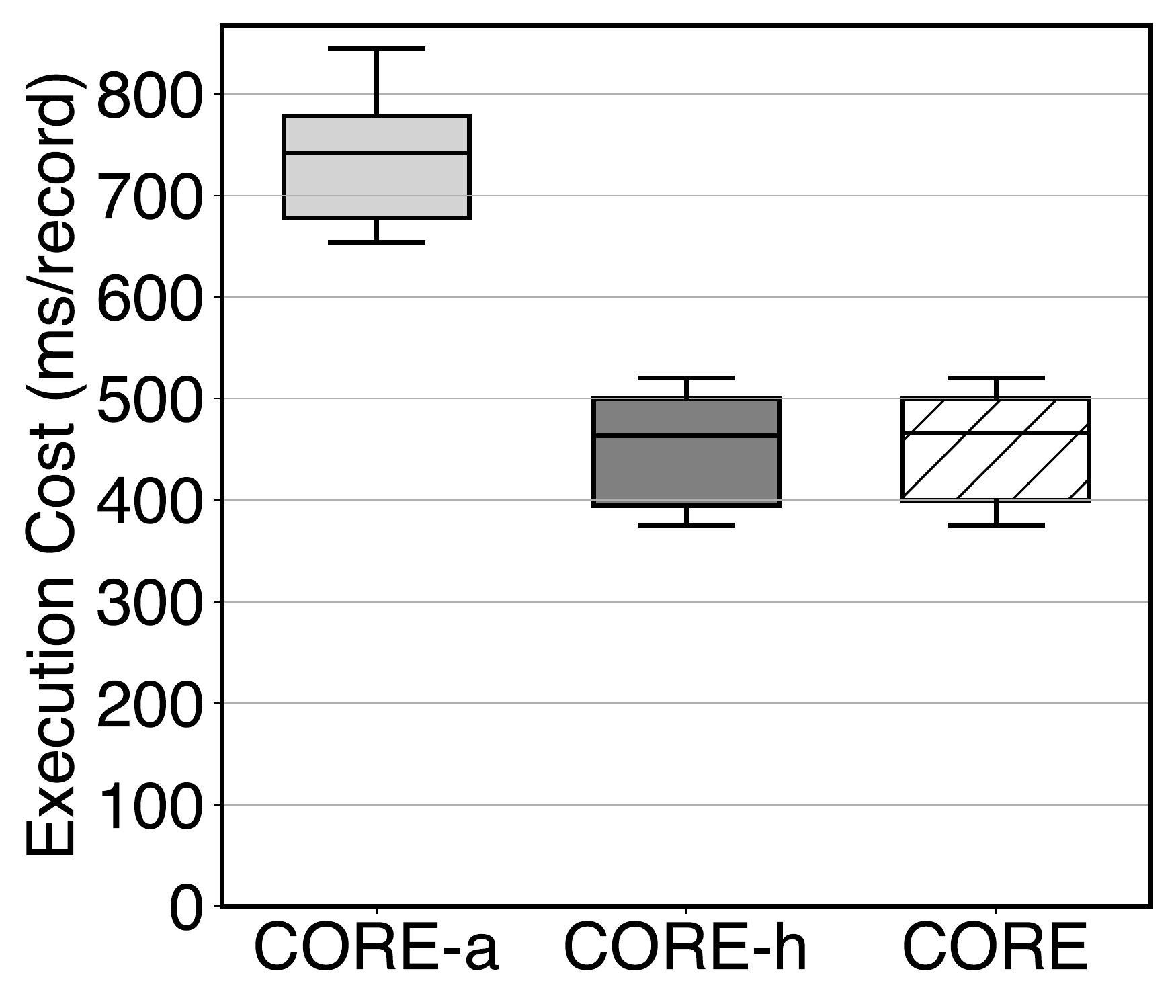}}
\subfloat[Optimization cost (UCF101).\label{fig:exp3optCostUcf101}]{\includegraphics[width=1.6in]{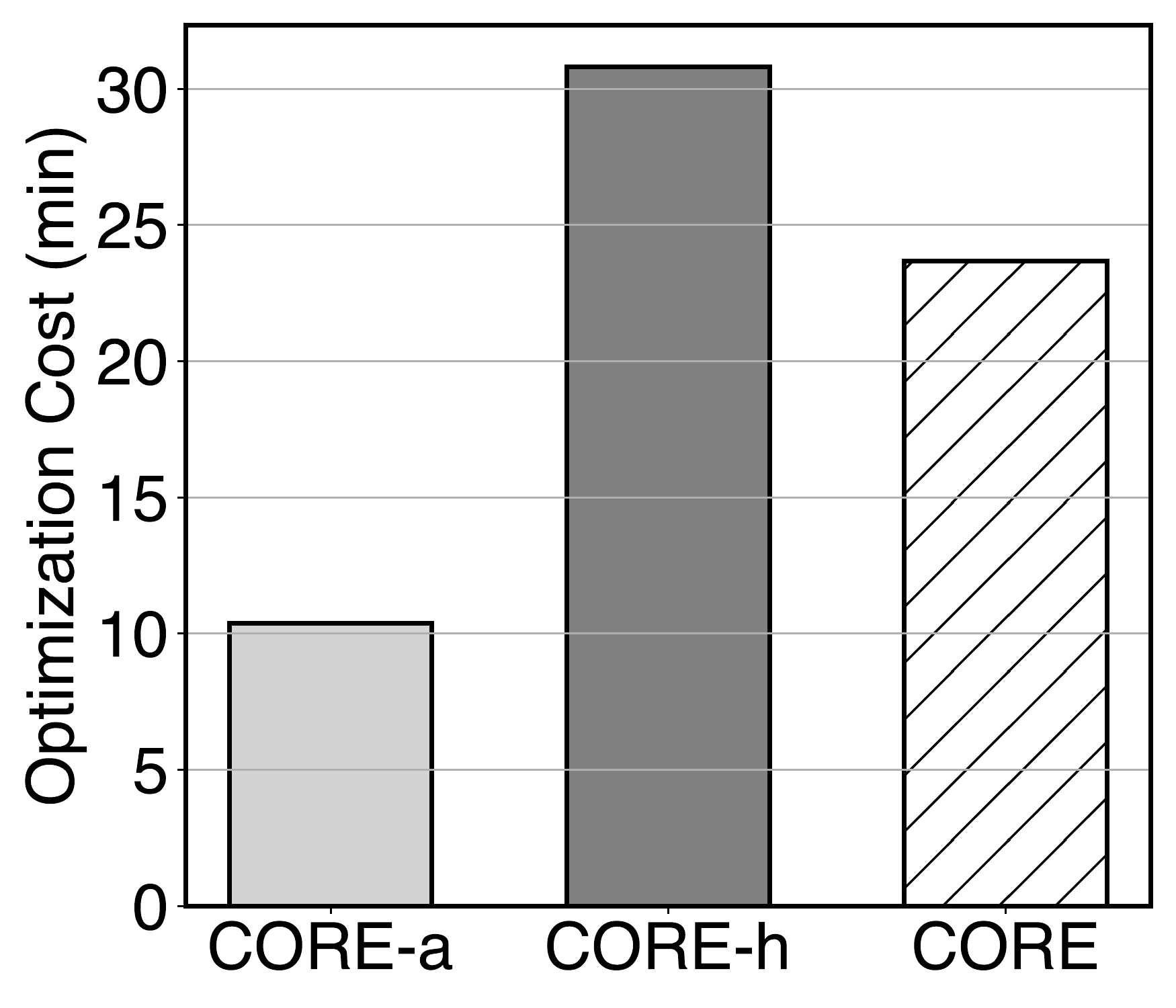}}
\caption{The execution costs and average optimization costs for queries over three datasets using {\name}, {\name}-a and {\name}-h.}
\label{fig:exp3exeOptCost3sets}
\end{figure}

We ran ten queries for each dataset using {\name}-a, {\name}-h, and {\name} with $\mathcal{A}=90\%$, and collected the execution costs for optimized plans and the average optimization costs to generate optimal plans. Figure~\ref{fig:exp3exeOptCost3sets} shows the results. We can see that {\name}-a had the worse execution cost compared to {\name} because {\name}-a did not use the optimal order. {\name} had similar execution costs to {\name}-h, but {\name}-h had much larger query optimization costs.  Table~\ref{tab:6detailcoresoptcost}, shows the average optimization cost including labeling, training, and searching using {\name}-a, {\name}-h, and {\name}. We can see that {\name} reduced the labeling, training and searching times compared to {\name}-h. This result indicated that the branch-and-bound search algorithm in {\name} successfully pruned some nodes in the tree and reduced the optimization overhead. In general, the branch-and-bound search algorithm found the optimal order. Therefore, both the Algorithm~\ref{alg:accuracy-allocation} for $\mathbb{A}$ and Algorithm~\ref{alg:B-B-pruning} for $\mathbb{H}$ successfully  accelerated the ML inference process.

\begin{table}[hbt!]
\small
\centering
\begin{tabular}{c|c|c|c|c|c}
\hline
& Labeling& Training & Searching& QO & QO\\
& Time& Time & Time& Time& Time\\
&(min) &(min) &(min) &(min) &pct.(\%)\\
\hline
{\name}-a & 1.37 & 0.15 & 1.78 & 3.30 & 0.38\\
{\name}-h & 6.51 & 0.57 & 4.69 & 11.78 & 1.74\\
{\name} & 1.84 & 0.44 & 2.61 & 4.91 & 0.70\\
\hline
\end{tabular}
\caption{Optimization costs of {\name} variants on the Twitter dataset.}
\label{tab:6detailcoresoptcost}
\end{table}

\subsection{Scalability} 

\begin{figure}[hbt!]
\subfloat[]{\includegraphics[width=1.6in]{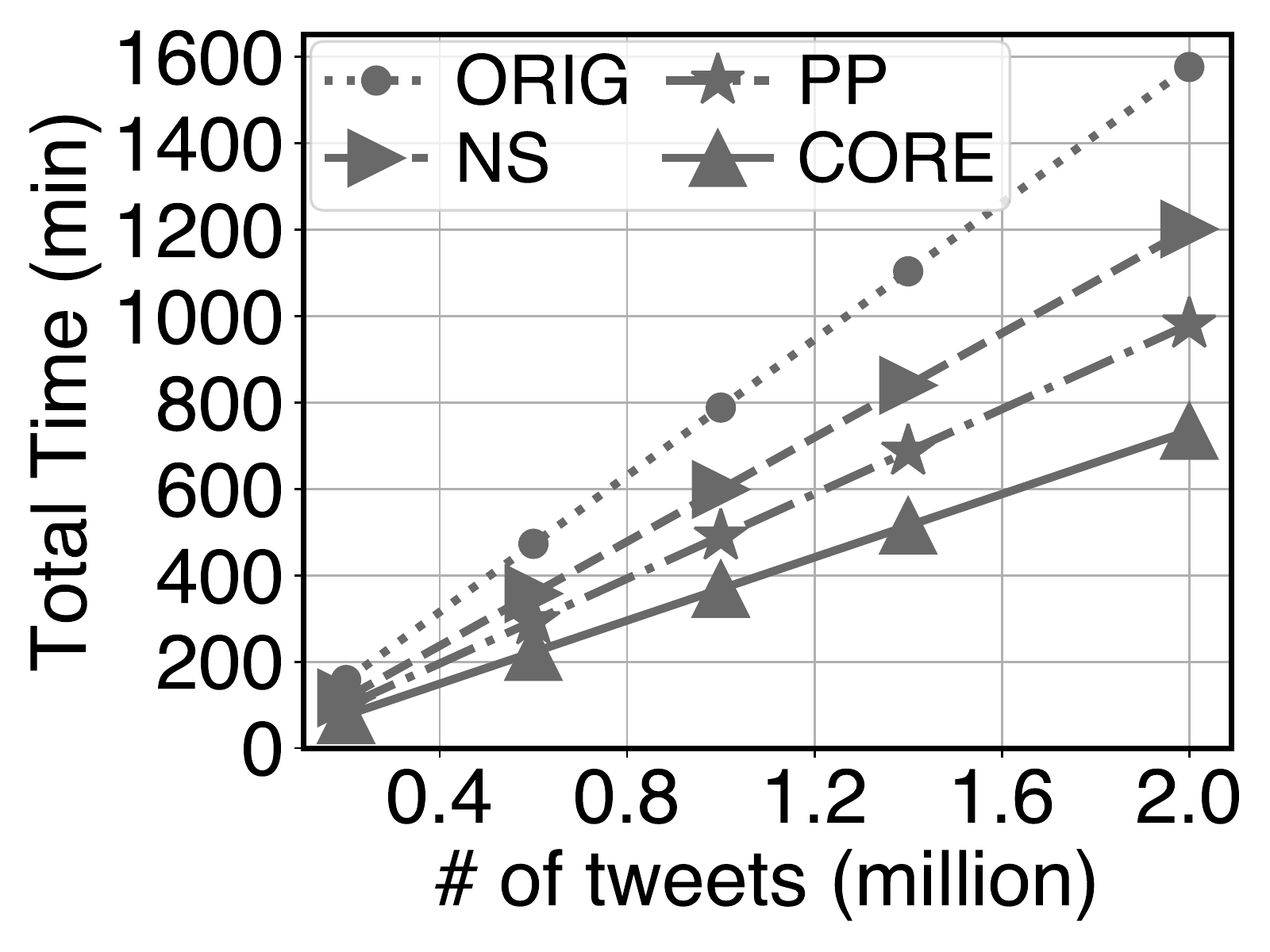}}
\subfloat[]{\includegraphics[width=1.6in]{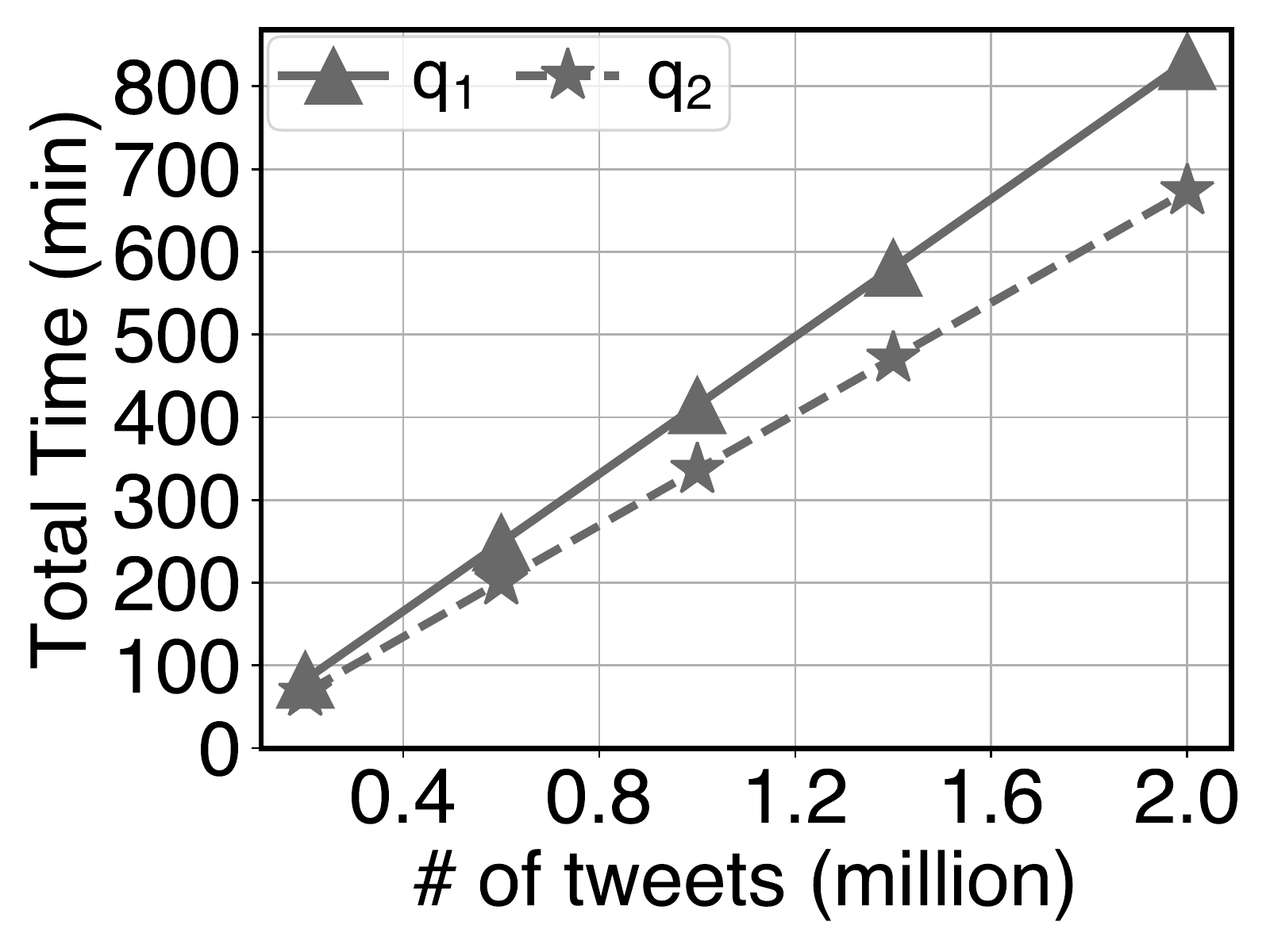}}
\caption{(a) The average total processing time (including optimization cost) using {\name}, {\sf ORIG}, {\sf NS}, and {\sf PP} on ten queries over the Twitter dataset with different input sizes. (b) The total times of two sample queries: $q_1$ and $q_2$, with different input sizes.}
\label{fig:exp2scaleDataSizeAve}
\end{figure}

We evaluated the scalability of {\name} by increasing the number of records in the Twitter dataset. We started with 0.2 million tweets and gradually increased the data size to 2 million tweets. We ran the ten queries with $\mathcal{A}=90\%$ using {\sf ORIG}, {\sf NS}, {\sf PP}, and  {\name}, and collected the total processing times at different data sizes. Figure~\ref{fig:exp2scaleDataSizeAve} shows the average total processing time using {\sf ORIG}, {\sf NS}, {\sf PP}, and  {\name}. We also presented the total times for two example queries using {\name} at different data sizes. The results show that  {\name} scaled up well, and outperformed the other three baseline approaches at all data sizes.

\subsection{Effect of Query Accuracy}
\label{sec:expSensitivity}

We evaluated the impact of the target accuracy $\mathcal{A}$ on {\name} by increasing $\mathcal{A}$. We started from $\mathcal{A}=90\%$, and linearly increased it to $\mathcal{A}=98\%$. We collected the execution costs of optimized plans for the ten queries over the Twitter dataset using {\sf ORIG}, {\sf NS}, {\sf PP}, and {\name} with different target accuracy values.

\begin{figure}[hbt!]
\subfloat[\label{fig:exp4exeCostAccuracyAve}]{\includegraphics[width=1.6in]{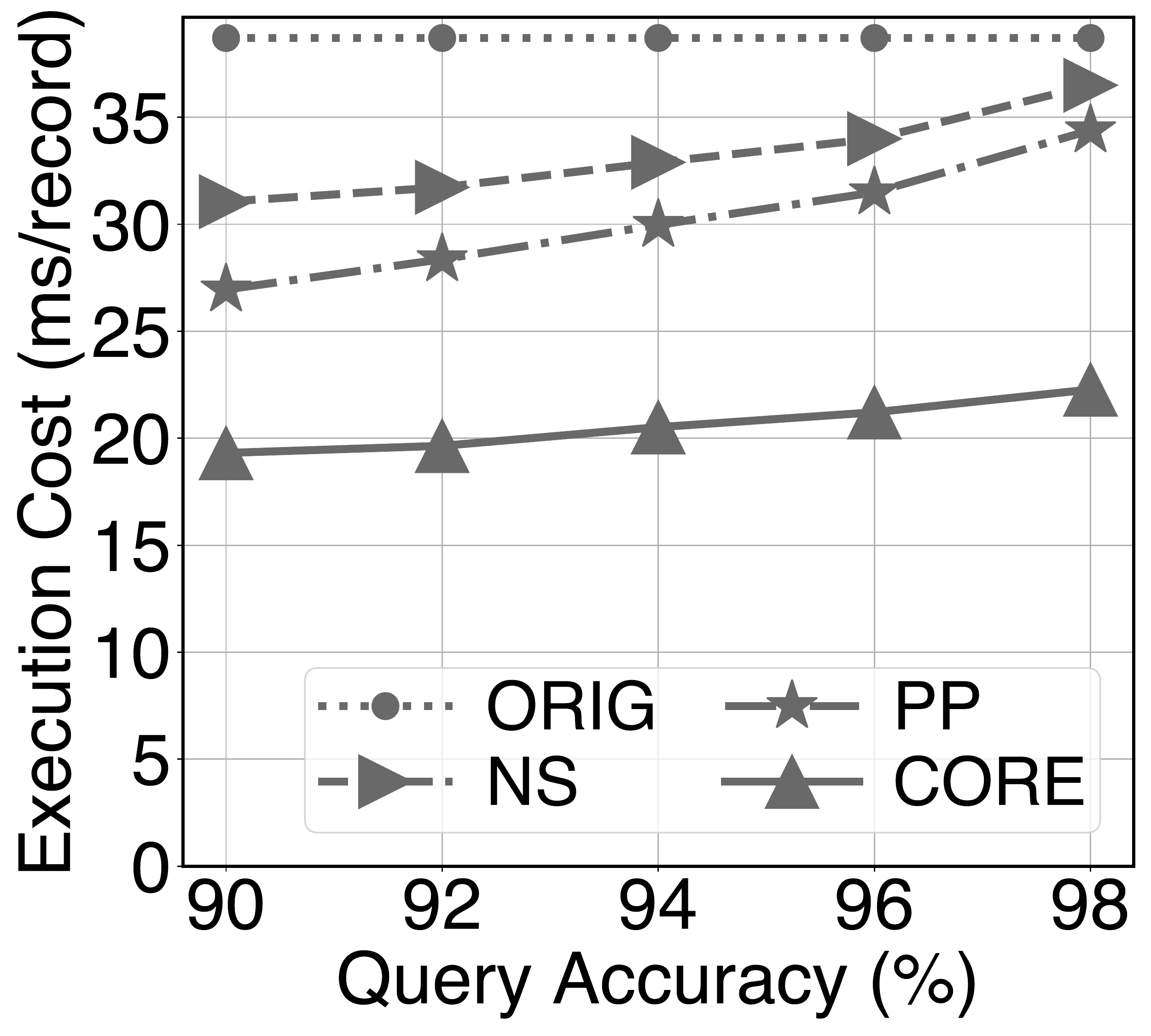}}
\subfloat[\label{fig:exp4exeCostAccuracy3queries}]{\includegraphics[width=1.6in]{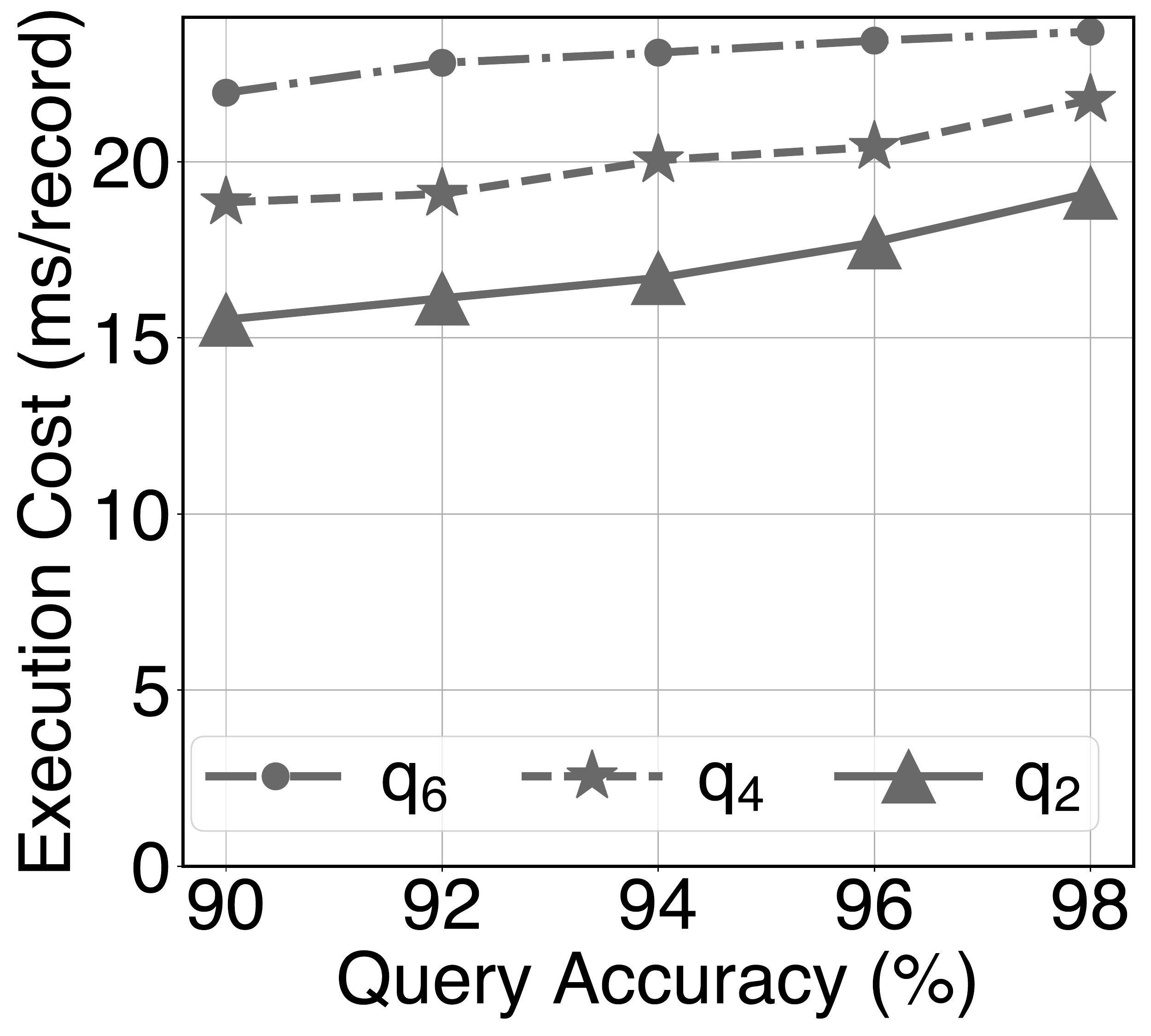}}
\caption{(a) The average execution costs of optimized plans for ten queries over the Twitter dataset with different $\mathcal{A}$ values. (b) The execution costs of three sample queries: $q_2$, $q_4$, and ${q_6}$, with different target accuracies.}
\label{fig:6exp4}
\end{figure}

Figure~\ref{fig:exp4exeCostAccuracyAve} shows the average execution costs for the ten queries using {\sf ORIG}, {\sf NS}, {\sf PP}, and {\name}. We also presented the execution costs for three example queries using {\name} with different target accuracy values in Figure~\ref{fig:exp4exeCostAccuracy3queries}. The results indicated that {\name} outperformed {\sf ORIG}, {\sf NS}, and {\sf PP} in different accuracy settings. Moreover, the execution costs increased for all the baselines when the target accuracy increased. In addition, Table~\ref{tab:6detailAccuracyOptCost} shows the percentage of the query optimization time relative to the total processing time in the same setting. Similar to the observations in Section~\ref{sec:6exp63}, the query optimization in {\name} with different accuracy targets still had a smaller overhead relative to the total processing time.

\begin{table}[hbt!]
\small
\centering
\begin{tabular}{|m{1.55cm}|m{0.9cm}|m{0.9cm}|m{0.9cm}|m{0.9cm}|m{0.9cm}|}
\hline
QO cost (min) / pct(\%) & ${\mathcal{A}=90\%}$ & ${\mathcal{A}=92\%}$ & ${\mathcal{A}=94\%}$ & ${\mathcal{A}=96\%}$ & ${\mathcal{A}=98\%}$\\
\hline
${q_2}$ & 1.50/0.11 & 1.54/0.11 & 1.50/0.11 & 1.48/0.11 & 1.48/0.11\\
\hline
${q_4}$ & 1.79/0.14 & 1.60/0.13 & 1.46/0.12& 1.52/0.12 & 1.50/0.12\\
\hline
${q_6}$ & 4.73/0.35 & 5.28/0.39 & 8.31/0.61 & 6.03/0.45 & 3.83/0.28\\
\hline
avg. & 4.57/0.36 & 4.83/0.38 & 5.07/0.40 & 4.30/0.34 & 3.24/0.25\\
\hline
\end{tabular}
\caption{The optimization costs for ${q_2, q_4}$ and ${q_6}$ with different target accuracy parameters (${\mathcal{A}}$). Each cell contains the QO costs in minutes and the percentage relative to the total query processing cost.}
\label{tab:6detailAccuracyOptCost}
\end{table}
\section{conclusions}
\label{sec:conclusions}

In this paper we proposed a novel query optimizer, {\name}, to accelerate ML inference queries. It improved state-of-the-art techniques by relaxing the assumption about independence of query predicates. {\name} uses a small computation overhead and leverages a branch-and-bound searching algorithm for finding an optimal order of proxy models with parameters for each predicate. It reduced the overhead by reusing intermediate results during the proxy model construction and pruning candidate plans. We conducted a thorough experimental evaluation and showed that {\name} significantly reduced the ML inference execution cost.


\balance


\bibliographystyle{abbrv}
\bibliography{bibliography}  



\end{document}